\documentclass[aps,prx,twocolumn,groupedaddress]{revtex4-1}

\usepackage{outlines}
\usepackage{dsfont}
\usepackage{enumitem}
\usepackage{braket}
\usepackage[tbtags]{amsmath}
\usepackage{mathtools}
\usepackage{amsfonts,amssymb,amsbsy}
\usepackage{graphicx}
\usepackage{MnSymbol}
\usepackage{xcolor}
\usepackage{hyperref}

\hypersetup{
    colorlinks=true,
    linkcolor=blue,
    filecolor=magenta,      
    urlcolor=red,
}

\begin{document}


\title{Subspace benchmarking high-fidelity entangling operations with trapped ions}


\author{C. H. Baldwin}
\email{charles.baldwin@honeywell.com}
\author{B. J. Bjork}
\author{J. P. Gaebler}
\author{D. Hayes}
\author{D. Stack}

\affiliation{Honeywell $|$ Quantum Solutions}

\date{\today}

\begin{abstract}
We present a new and simplified two-qubit randomized benchmarking procedure that operates only in the symmetric
subspace of a pair of qubits and is well suited for benchmarking trapped-ion systems. By performing benchmarking only in the symmetric subspace, we 
drastically reduce the experimental complexity, number of gates required, and run time. The protocol is
demonstrated on trapped ions using collective single-qubit rotations and the M\o lmer-S\o renson (MS) interaction to
estimate an entangling gate error of $2(1) \times 10^{-3}$. We analyze the expected errors in the MS gate and
find that population remains mostly in the symmetric subspace. The errors that mix symmetric and anti-symmetric
subspaces appear as leakage and we characterize them by combining our protocol
with recently proposed leakage benchmarking. Generalizations and limitations of the protocol are also discussed.
\end{abstract}


\maketitle

\section{Introduction}
Like any complex machine, the construction of a large-scale quantum computer will require a rapid design-and-test cycle of individual components and their assemblage. Early recognition of this need by quantum information scientists led to widespread, (but not universal~\cite{Nielsen05}), adoption of the process fidelity as the gold-standard for the quality of quantum operations \cite{Wineland98}. While the quantity had solid theoretical motivations, experimentalists lacked a simple method for its measurement for two or more qubits. In practice, experimentalists were stuck choosing between comprehensive but resource-intensive process tomography~\cite{OBrien04} and convenient but less descriptive Bell-state tomography~\cite{Ballance16, Gaebler16, Benhelm08, Wright19}. 

The advent of randomized benchmarking (RB) marked a powerful advancement in experimental validation of quantum operations~\cite{Emerson05,Knill08}. The new tool provided algorithmic advantages, including robustness to state-preparation and measurement (SPAM) errors~\cite{Magesan11, Magesan12a},
multiplicative precision in estimates~\cite{Harper19}, and, under certain assumptions, an accurate direct estimate of the fidelity~\cite{Wallman18}. In the case of multi-qubit gate benchmarking, one disadvantage of the protocol is the requirement for individual addressing. While this demand is also needed for universal quantum computation, it can impose a significant engineering requirement on the design-and-test cycle, especially in trapped-ion systems. In this work, we alleviate this requirement and provide a benchmarking protocol that uses only global operations.

Other methods for RB with limited gate sets have also been considered in the context of logical RB~\cite{Brown18, Hashagen2018}, and as a way to simplify standard RB~\cite{Franca18, Helsen19b}, but these methods suffer from added sampling overhead and require a detailed study of the gates' group structure to identify different decay rates. Our method circumvents these drawbacks.

The basic idea behind our procedure is to isolate a subspace of the total system to perform RB, hence we call the method subspace randomized benchmarking (SRB). We then use leakage benchmarking~\cite{Wallman16, Wood18, Andrews19} to identify errors that may drive population outside of this subspace. We show that these different methods can be combined into a single procedure and each effect can be individually measured. This analysis can identify many common errors, and we consider several examples numerically.

In addition to developing the theory behind SRB, we demonstrate the procedure in a trapped ion system. We benchmark the M\o lmer-S\o renson (MS) interaction and collective single-qubit gates (identical single-qubit gates applied to all ions) using two ions in a four-ion crystal. The other ions are a different species and are used for sympathetic cooling to prevent the temperature from changing throughout the gate sequences. With this system, we are able to run gate sequences with as many as 400 MS gates. The SRB procedure returns an estimated entangling gate error of $2(1) \times 10^{-3}$.

The remainder of this paper is organized into the following sections:
\begin{itemize}
	\item[\textbf{II.}] \textbf{Standard RB -} A review of the standard randomized benchmarking protocol and how to interpret the experimental results.
	\item[\textbf{III.}] \textbf{Subspace RB -} Our construction of SRB and how subspace leakage is considered.
	\item[\textbf{IV.}] \textbf{Extracting information from SRB -} Analysis of how different errors are manifested in SRB.
	\item[\textbf{V.}] \textbf{Experimental platform -} A description of the SRB demonstration.
	\item[\textbf{VI.}] \textbf{Results and discussion -} Estimates of the subspace and leakage errors.
	\item[\textbf{VII.}] \textbf{Generalized subspace RB -} Extensions to other systems.
\end{itemize}

\section{Standard Randomized Benchmarking} \label{sec:standard_RB}
We begin by outlining the standard RB procedure~\cite{Magesan12a} with further details given in App.~\ref{app:standard_RB}. RB provides an estimate of the fidelity of a finite set of gates
$\mathsf{C}=\{C_1, \dots, C_N\}$. The standard RB construction requires that 
the gates form a representation of a finite group, which is usually the Clifford group. A Clifford gate is the unitary implementation of a Clifford group element and defined by their action on Pauli gates. A Pauli gate is the unitary implementation of a Pauli group element, which we label  $\mathds{1}, X, Y$, and $Z$ and tensor products
thereof. A Clifford gate maps Pauli gates to Pauli gates
with a sign, e.g. $C X C^{\dagger} = -Y$. The Clifford gates are selected in RB for two reasons: (1) it is
efficient to simulate their evolution by the Gottesman-Knill theorem~\cite{Gottesman98}, and (2) they form a unitary two-design which simplifies the analysis of an RB experiment~\cite{Magesan12a}.

\begin{figure} 
  \centering
    \includegraphics[width=0.5\textwidth]{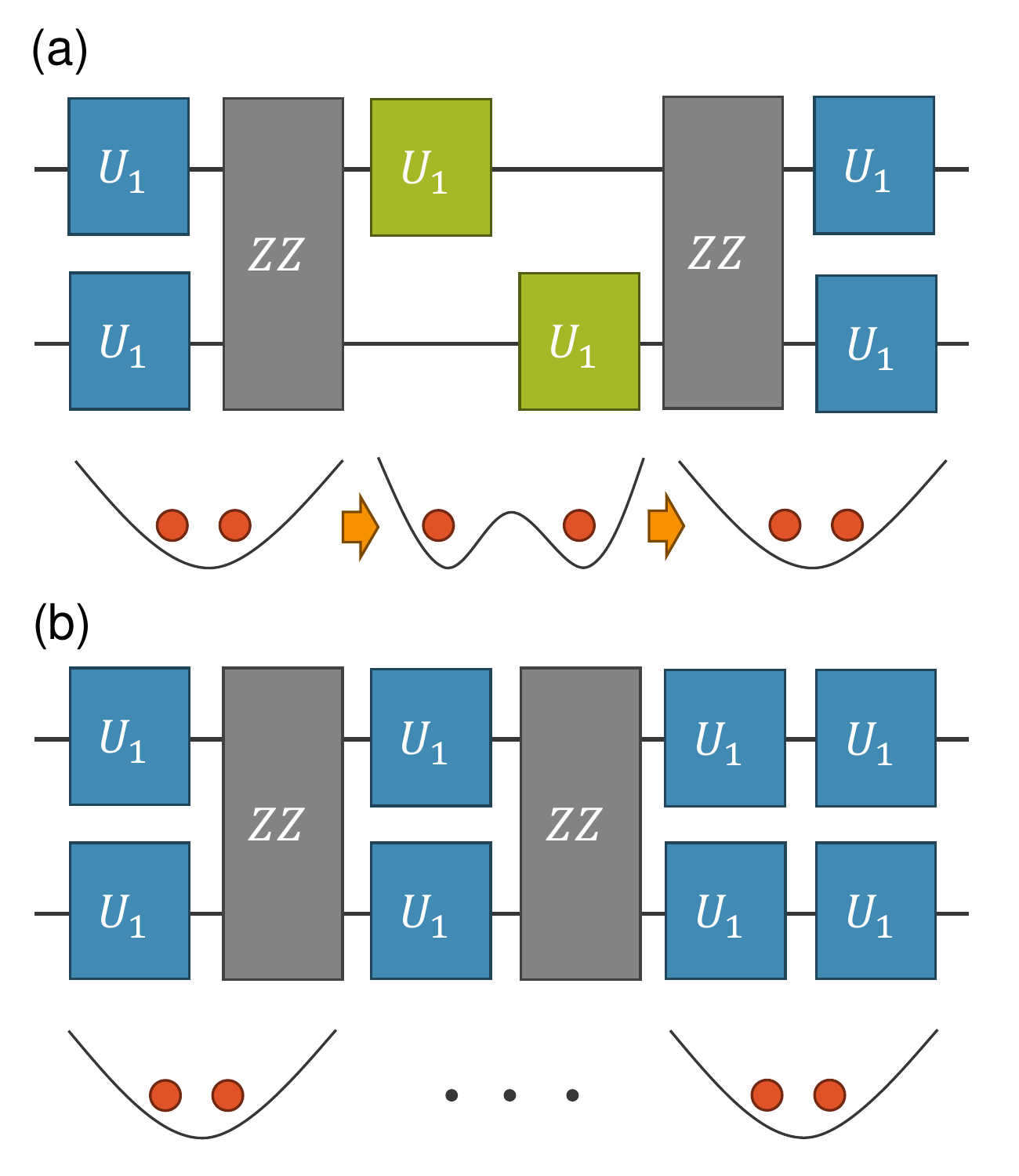}
\caption{Circuits used for standard RB and SRB. Blue gates are collective single-qubit gates, gray gates are entangling phase gates, and green gates are individually addressed single-qubit gates.  (a) Standard RB for a transport-based trapped-ion architecture. Individually addressed gates require a transport operation and additional cooling, which involves significant engineering overhead and may increase the error rate per Clifford. Alternative individual addressing methods exist but also have engineering overheads and similarly affect the error rates. (b) SRB only has collective and phase gates that do not require transport and cooling can be implemented during the collective gates. This reduces total run time and other errors.}
\label{fig:seq}
\end{figure}

A standard RB experiment has sequences of gates shown in Fig.~\ref{fig:seq}, and proceeds as follows:
\begin{enumerate}
\item Prepare $\ket{0}$
\item Randomly select $\ell$ gates from $\mathsf{C}$ and apply to $\ket{0}$
\item Calculate the inverting gate $C_{\textrm{inv}}= C^{\dagger}_1 \cdots C^{\dagger}_{\ell} \in \mathsf{C}$
\item Measure survival $p({\ell}) =|\braket{0 | C_{\textrm{inv}} C_{\ell} \cdots C_{1} | 0 }|^2$
\item Repeat (1-4) $K$ times with different random gates and estimate the average survival
$\overline{p}({\ell}) = \tfrac{1}{K} \sum_i p_{i}(\ell)$
\item Repeat (1-5) with different $\ell$ 
\end{enumerate}
We typically pick $\ell$ exponentially distributed between 1 and $\ell_{\textrm{max}}\approx1/\epsilon$ where $\epsilon$ is the 
expected error~\cite{Meier06}. We typically pick $K\approx 100$ sequences, but rigorous estimates also exist based on expected fidelity~\cite{Wallman14, Helsen19a}.

Although the net action is chosen to be the identity operation, noise and errors in the state-preparation, gates, and measurement cause $p(\ell) \leq 1$. The power of RB is to relate the
average survival $\overline{p}({\ell})$ to the process fidelity of the Clifford gates. If each Clifford gate has the same error channel then randomization over the Clifford gates depolarizes each error channel due to the unitary two-design property~\cite{Magesan11}. The resulting average survival is
\begin{equation} \label{eq:rb_decay}
\overline{p}({\ell}) =A r^{\ell} +B,
\end{equation}
where $A$ and $B$ relate to SPAM errors, and $r$ is referred to as the standard decay rate. By evaluating $\overline{p}(\ell)$
for different values of $\ell$, we can extract $A$, $B$, and $r$ by least-squares fitting. The process fidelity of the Clifford gate's error channel is then defined as
\begin{equation}
F_{\textrm{Cliff}} = \tfrac{(d^2-1) r + 1}{d^2}.
\end{equation}

In practice, Pauli and Clifford gates are implemented by compiling physically available gates. In trapped-ion systems (and many
other quantum systems) the physical single-qubit gates are implemented as rotations in the $X-Y$ plane of the Bloch sphere with
arbitrary amplitude $U(\theta,\phi) = \exp[-i \tfrac{\theta}{2}(X \cos(\phi) + Y \sin(\phi))]$. The entangling 
operations for many trapped-ion experiments, including ours, is the MS gate, $U_{MS}(\theta, \phi) = \exp[-i
\tfrac{\theta}{2} (X \cos(\phi)+ Y \sin(\phi))^{\otimes 2}]$. As discussed later, we use the phase gate
$U_{ZZ} = \exp[-i \tfrac{\pi}{4} ZZ]$, which is created by applying single-qubit ``wrapper pulses'' before and after the
MS gate~\cite{Lee05}. In the system considered, single-qubit gates are not individually addressable, but we do have access to collective single-qubit gates $U(\theta,\phi) = \exp[-i \tfrac{\theta}{2}(X \cos(\phi) + Y \sin(\phi))]^{\otimes 2}$.

\section{Subspace randomized benchmarking} \label{sec:ssrb}
Our procedure for RB, which we call subspace randomized benchmarking (SRB), follows the same general prescription
outlined above except we consider a restricted set of physical gates. This restriction could be due to the physical limitations,
i.e. the system cannot implement a universal set, or because some gates have much lower fidelity and must be
avoided. In either case, the remaining set of gates is not universal, rendering standard RB impossible. We now construct a protocol with a restricted set of gates that retains most of RB's favorable attributes and still yields
detailed information about the gate in question.

We introduce the procedure in the context of a trapped-ion system where the two qubits are encoded in two ions in the same potential well.
Without tightly focused laser beams or complex transport operations, the available one- and two-qubit gates will be symmetric 
with respect to qubit exchange. This limited set of physical gates cannot generate the two-qubit Clifford gates needed for 
standard RB. Nevertheless, the goal of the setup is to determine a quantitative measure of the
entangling gate's performance. This is a common setup for many trapped-ion experiments that are trying to demonstrate
new entangling gates or high-fidelity performance~\cite{Ballance16, Gaebler16, Wright19}.

The first step in SRB is to identify a subspace of the total Hilbert space where we can perform RB, which we call the RB
subspace. For the trapped-ion system, the symmetric subspace of the two-qubits is a natural choice. States in this subspace are symmetric under the
SWAP operation, with basis $\{\ket{00},\tfrac{1}{\sqrt{2}}(\ket{01} + \ket{10}), \ket{11} \}$. The remaining space is spanned by the anti-symmetric state $\tfrac{1}{\sqrt{2}}(\ket{01} - \ket{10})$, and we refer to it as
the leakage subspace. It is easy to prepare the $\ket{00}$ state via optical pumping with high fidelity~\cite{Olmschenk07}. The
collective single-qubit gates and phase gate are block-diagonal in this basis, and therefore maintain population in each
subspace. Measurements in the system return the number of qubits in the $\ket{1}$ state (the number of ions fluorescing). If the population is isolated to the symmetric subspace then this measurement
corresponds to a projection measurement in the basis for RB subspace defined above.

The next step is to identify a set of gates in the RB subspace that form a unitary two-design, which was the property that allowed us to derive Eq.~\eqref{eq:rb_decay}. In standard RB, we use Clifford gates, which map Pauli gates to Pauli gates. For prime dimensions, we can generalize the Pauli gates to Weyl gates and the corresponding Clifford gates map Weyl gates to Weyl gates~\cite{Hostens05}. These generalized Cliffords also form a unitary-two design, which will act to depolarize any error channel~\cite{Dankert09}. For $d_1 = 3$ (or qutrit) there are 9 qutrit-Weyl gates and 216 qutrit-Clifford gates. 

To generate each qutrit-Clifford gate we use techniques from optimal control theory. In the RB subspace the collective
gates act as spin-1 $J_x$ and $J_y$ angular momentum operators, and the phase gate acts as a $J_z^2$ operator. Then,
control of the symmetric subspace is analogous to control of a spin-1 particle. We make use of the Cartan decomposition~\cite{Khaneja01,Shauro2015} to
decide the order of collective and phase gates and then perform a numerical search~\cite{Machnes11} for each of the 216 qutrit-Clifford gates to
find amplitudes and phases. We find that only three phase gates are required per qutrit-Clifford arranged as shown in Fig.~\ref{fig:seq}. This does require full simulation of the system, but for two qubits, we find that it takes a reasonable amount of time. Further details are given in App.~\ref{app:control}. 

The protocol then proceeds the same as standard RB but with the qutrit-Clifford gates. After taking data with various sequence lengths, we estimate the RB decay parameter $r$, and calculate the subspace fidelity, which we denote with lower-case $f$ to differentiate from the full-space fidelity $F$,
\begin{equation}
f_{\textrm{Cliff}} = \tfrac{(d_1^2-1) r + 1}{d_1^2},
\end{equation}
where $d_1$ is the dimension of the RB subspace. The subspace fidelity only provides an estimate for errors that are isolated to the RB subspace. We call the procedure given so far SRB-\textit{lite}.

An obvious problem with SRB-\textit{lite} is that there may exist errors that are not isolated to the RB subspace. In a quantum algorithm, the input state to each gate is unlikely to be symmetric and only measuring errors on the symmetric subspace may miss errors that degrade the computation. Moreover, some errors
might couple the RB and leakage subspaces, violating the assumptions we made to derive the RB decay curve.
A similar problem arises in single-qubit RB if there are errors that cause population to leave the 
computational subspace, called leakage errors. Previously, several protocols to measure leakage rates were proposed, which we refer to as leakage benchmarking (LB)~\cite{Wallman16, Wood18, Andrews19}. We now show LB can be performed along with SRB-\textit{lite} to create a more complete picture of the errors in the MS gate. We call this combined procedure SRB.

In Ref.~\cite{Wood18}, leakage errors are quantified by a leakage rate $L$ (rate that which population moves from RB to leakage subspace) and seepage rate $S$ (rate that which population moves from the leakage to the RB subspace),
\begin{align}
L &= \tfrac{1}{\sqrt{d_1 d_2}} \textrm{Tr}(\Lambda[\mathds{1}_1]\mathds{1}_2), \\
S &= \tfrac{1}{\sqrt{d_1 d_2}} \textrm{Tr}(\mathds{1}_1 \Lambda[\mathds{1}_2]),
\end{align}
where $\mathds{1}_1$ and $\mathds{1}_2$ are projectors onto the RB and leakage subspaces respectively. Our definitions differ slightly from
Ref.~\cite{Wood18} by factors $d_1$ (dimension of RB subspace) and $d_2$ (dimension of the leakage subspace) in order to simplify the derivations in App.~\ref{app:ssrb}.
Leakage and seepage errors cause additional terms to enter into Eq.~\eqref{eq:rb_decay}~\cite{Wallman16, Wood18},
\begin{equation} \label{eq:rb_wleakage}
\overline{p}({\ell})=A r^{\ell} + B + C t^{\ell+1},
\end{equation}
where $A$, $B$ and $C$ relate to SPAM errors, $r$ is the standard RB decay, and $t$ relates to leakage and seepage rates. 

We use a different technique that more effectively extracts the decay rates $r$ and $t$ than previous LB proposals.  It requires including an additional random Weyl gate $Q_k$, which can be
compiled into the final gate $C_{\textrm{inv}} = Q_k C_{\ell}^{\dagger} \dots C_1^{\dagger}$. This method was also proposed in Refs.~\cite{Knill08, Meier06} to catch pathological errors in standard RB. A similar method was explored in Ref.~\cite{Harper19} to reduce the fitting parameters in order to derive multiplicative precision in standard RB. In addition to both of these advantages, we find that it also allows us to separate the analysis of the standard decay $r$ from the effects of leakage errors and reduce the assumptions required in the derivation. Further details are given in App.~\ref{app:ssrb}.

We again randomize over all sequences, which includes randomization over the final Weyl gate. Then, we process the
data in two ways:
\begin{itemize}
\item \textit{Standard analysis}: For the standard RB decay, we only use the measurement outcomes $m_k$ such that $|\braket{m_k | Q_k | 0}|^2= 1$ in the absence of errors. Averaging over Weyl gates, with the related outcome $m_k$, fixes $B = 1/d_1$ and $C =0$, leaving a single exponential
decay to fit, 
\begin{equation} \label{eq:standard_decay}
\overline{p}_S({\ell}) =A r^{\ell} + \tfrac{1}{d_1},
\end{equation}
This differs from Ref.~\cite{Wood18} since $B$ is fixed but does not require assumptions on SPAM errors like in Ref.~\cite{Andrews19}. 

\item \textit{Leakage analysis}: To measure the leakage decay, we pick $m=0$ outcomes, corresponding to the survival with the
additional Weyl gate. Averaging over all sequences gives the average leakage decay
\begin{equation} \label{eq:leakage_decay}
\overline{p}_L({\ell}) = C t^{\ell+1} + B.
\end{equation}
\end{itemize}

In Ref.~\cite{Wood18} $L$ and $S$ are used to quantify leakage errors, but we propose that $t$ is more reliable and almost as
valuable. This is because the estimates of $L$ and $S$ rely on the fits of $C$ and $B$, which are limited by finite sampling and SPAM errors. The estimation method for $t$ is almost identical to the estimation of $r$, and so follows the standard RB paradigm. The downside of only studying $t$ is that we cannot differentiate between leakage and seepage errors, but our analysis finds that these rates are usually similar for standard errors in the trapped-ion system.

\begin{figure} 
  \centering
    \includegraphics[width=0.5\textwidth]{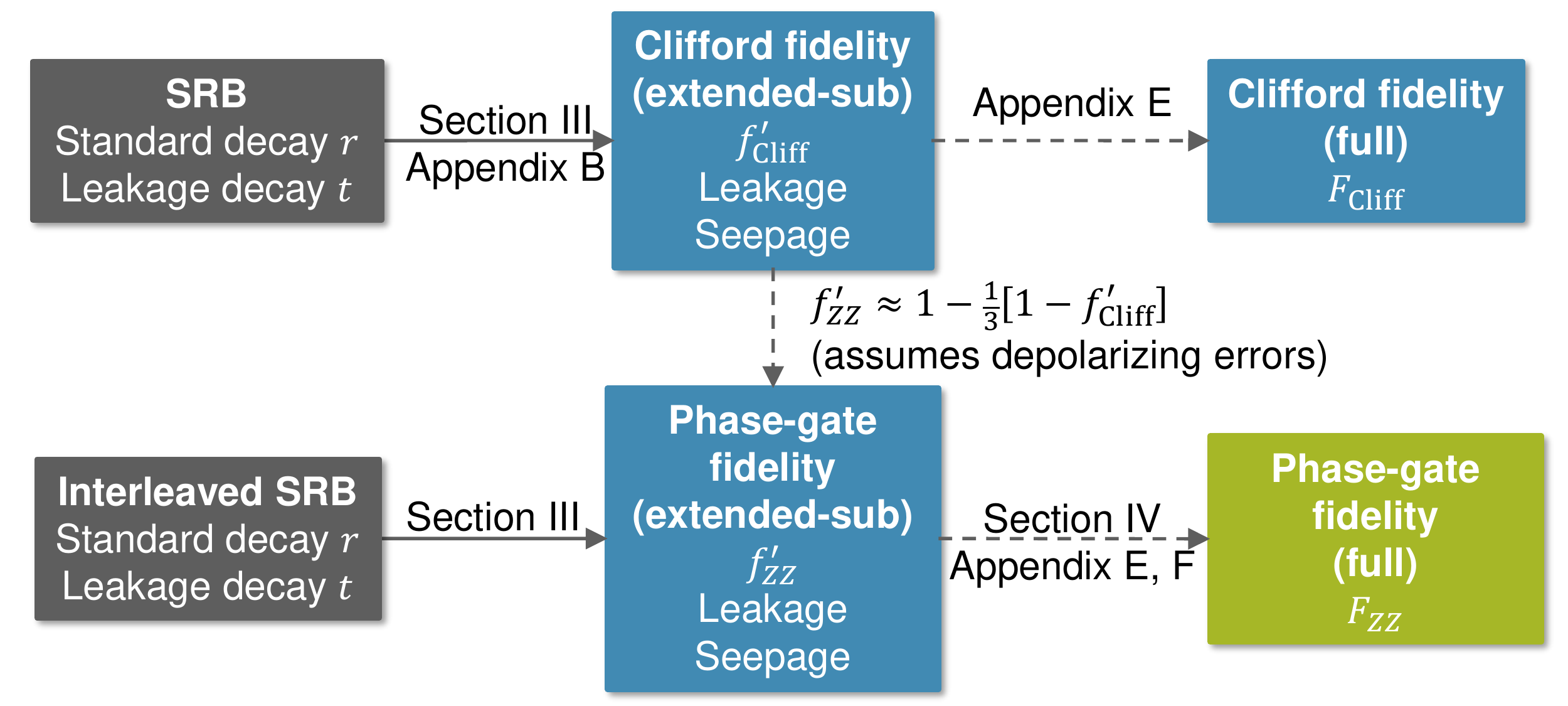}
\caption{Connections between measured quantities in SRB (standard decay rate $r$ and leakage decay rate $t$), the
extended-sub fidelity, and the full fidelity. The top row shows the
connections between SRB and qutrit-Clifford gates. The bottom row shows the connection between interleaved SRB and the
phase gate. If we assume the phase gate errors are depolarizing and dominate each Clifford gate then we can connect the top
and bottom rows by the approximation $f_{ZZ}' = 1- \tfrac{1}{3}(1- f_{\textrm{Cliff}}')$.}
\label{fig:connections}
\end{figure}

The results of both analyses then contribute to an estimate of the subspace fidelity of the qutrit-Clifford gates on the RB subspace.
The leakage rate contributes to this estimate since the leakage errors reduce population in the RB subspace. We propose a variant of subspace fidelity that includes the seepage rate 
and is only dependent on $r$, $t$, and the dimensions, which we call the extended-sub fidelity,
\begin{equation} \label{eq:extnd_fid}
f_{\textrm{Cliff}}' = \tfrac{d_1^2}{d_1^2 + 1}f_{\textrm{Cliff}} + \tfrac{1}{d_1^2 + 1} S=\tfrac{8 r + t + 1}{10},
\end{equation}
where the second equality is specific for the trapped-ion system. This quantity has the similar advantages of estimating
$t$ over $L$ and $S$ as well as capturing additional information about seepage errors in the fidelity. 

To extract the extended-sub fidelity of a single phase gate we scale $f_{\textrm{Cliff}}'$ by the number of phase gates per Clifford, which is three,
\begin{equation}
f_{ZZ}' \approx 1-\tfrac{1}{3}(1-f_{\textrm{Cliff}}') = \frac{8(r/3) + t/3 + 7}{10}
\end{equation}
A similar procedure can be applied with $f_{\textrm{Cliff}}$ from SRB-\textit{lite}. This procedure is only exact when phase-gate errors dominate and each phase-gate has a depolarizing error channel. It is very likely that the phase gate errors dominate but unlikely that the errors are purely depolarizing in any real experiment. The quantity still serves as a rough estimate of the contribution of each phase gate to the error in the qutrit-Clifford gates. We report the estimated phase-gate error, which is $1-f_{ZZ}'$.

Another option to estimate phase gate fidelity is to perform interleaved randomized benchmarking (IRB)~\cite{Magesan12b} to bound the phase-gate error. This is shown as the bottom row of Fig.~\ref{fig:connections}. For SRB, such an interleaved procedure is more complicated since the phase gate is not a qutrit-Clifford gate. Instead of randomizing over the qutrit-Clifford gates we randomize over the larger group of arbitrary SU(3) unitary operators. The procedure is then the same as standard IRB; perform SRB and then perform SRB with interleaved phase gates. The standard IRB procedure also assumes that the errors are depolarizing and returns weak bounds on the fidelity if this is not true~\cite{Magesan12b}. While this could be improved by including other experiments (like unitarity benchmarking~\cite{Carignan19}), we opt for the simpler procedure given above for estimating the phase-gate error and stress that the estimates are subject to the depolarizing assumption.

\section{Extracting information from SRB} \label{sec:sims}
To understand how errors affect SRB, we preformed an extensive study of typical errors in trapped-ion experiments. We found that each
error falls into one of two categories: (1) Errors that preserve the subspaces and (2) errors that mix population
between RB and leakage subspaces. The results are summarized in Table~\ref{tab:table1} and further details of each error channel are given in App.~\ref{app:errors}.

Table~\ref{tab:table1} also shows how errors have different characteristics that are measurable by SRB. For example,
intensity errors fall into type-1 since they preserve the subspaces, and therefore we expect an SRB experiment to
measure $r< 1$ and $t \approx 1$. Spontaneous emission causes mixing between the subspaces so that we expect $r< 1$ and $t < 1$. An interesting result from the study is that several errors can be tuned between type-1 and -2 by adjusting ion
spacing and the laser/motional phase. This could also be used to further isolate dominant errors in the phase gate.

While SRB can detect most of the common errors in the phase gate, it will miss one type of error. We refer to these as `worst-case' errors, and they correspond to errors that change the relative phase between the RB and leakage subspaces. For example, if the ions reordered this would look like a logical SWAP error, which generates a $-1$ phase between the two subspaces. This type of error will go unnoticed in an SRB experiment but could cause serious problems in a full computation. In general though, these errors are unlikely as they do not arise from any of the known MS error mechanisms~\cite{Sorensen00} and our idling SWAP rate, (from background gas collisions), is observed to be on the timescale of tens of minutes.
\begin{table}
  \begin{center}
    \begin{tabular}{|l|c|c|c|c|}
      \hline
\textrm{\textbf{Error channel}$\quad\quad\quad\quad\quad$} & $\,\mathbf{m}_+,\mathbf{l}_+\,$ & $\,\mathbf{m}_+,\mathbf{l}_-\,$ & $\, \mathbf{m}_-,\mathbf{l}_+\,$ &
$\,\mathbf{m}_-,\mathbf{l}_-\,$ \\
      \hline
      Intensity error & $\checkmark$ & $\checkmark$ & $\checkmark$ & $\checkmark$ \\
      \hline
      Lamb-Dicke & $\checkmark$ & $\checkmark$ & $\checkmark$ & $\checkmark$\\
      \hline
      Debye-Waller & $\checkmark$ & $\checkmark$ &  $\checkmark$ & $\checkmark$\\
      \hline
Trap frequency error & $\checkmark$ & $\rightleftarrows$ & $\rightleftarrows$ & $\checkmark$\\      \hline
Spectator mode coupling & $\rightleftarrows$ & $\rightleftarrows$ & $\rightleftarrows$ &
$\rightleftarrows$\\
      \hline
      Carrier coupling & $\checkmark$ &$\rightleftarrows$ & $\checkmark$ & $\rightleftarrows$\\
      \hline
      Carrier frequency error & $\checkmark$ & $\checkmark$ & $\checkmark$ & $\checkmark$ \\
      \hline
      Uneven ion illumination & $\rightleftarrows$ &  $\rightleftarrows$ & $\rightleftarrows$ & $\rightleftarrows$\\
      \hline
      Heating & $\checkmark$ & $\checkmark$ & $\checkmark$ & $\checkmark$\\
      \hline
	Spontaneous emission & $\rightleftarrows$ & $\rightleftarrows$ & $\rightleftarrows$ &
$\rightleftarrows$\\	\hline
	Stray optical pumping & $\rightleftarrows$ & $\rightleftarrows$ & $\rightleftarrows$ &
$\rightleftarrows$\\ \hline
	Inhomogeneous fields & $\rightleftarrows$ & $\rightleftarrows$ & $\rightleftarrows$ &
$\rightleftarrows$\\
      \hline
    \end{tabular}
\caption{This table categorizes common errors in the phase gate according to whether or
not they preserve the RB and leakage subspaces. The different columns denote the qubit exchange symmetry
of the motion (m) and that of the laser phase (l).
Errors that preserve the subspaces are marked $\checkmark$ (type-1), whereas the errors that mix the
subspaces are marked $\rightleftarrows$ (type-2).}
\label{tab:table1}
  \end{center}
\end{table}

\section{Experimental platform}
We implement our SRB procedure in a linear RF Paul trap with a chain of two $^{171}$Yb$^{+}$ data qubits and two $^{138}
$Ba$^{+}$ ions for sympathetic cooling. The motion of the ions is described by shared motional modes with
transverse (X,Y) and axial (Z) harmonic confinement at frequencies $\{\omega_x,\omega_y,\omega_z\} = 2\pi \times
\{2.72,3.24,1.15\}$ MHz for a single $^{171}$Yb$^{+}$. The axial modes of motion and the three lowest frequency radial modes of
the four ion chain are cooled to $\bar{n} < 1$ via resolved sideband cooling of both Yb$^+$ and Ba$^+$ before the start
of an SRB sequence~\cite{Monroe95}. The axial in-phase mode is cooled during an SRB sequence using sideband cooling on the Ba
ions to mitigate effects of motional heating that takes place during the SRB sequence.

The data qubits are encoded in the $^2 S_{1/2}$ hyperfine levels of the $^{171}$Yb$^+$ and
initialized and read out using standard techniques~\cite{Olmschenk07}. The qubit levels are coupled via Raman transitions
driven by a pair of $375$ nm phase-locked lasers with a $12.6$ GHz frequency offset to match the
hyperfine frequency $\omega_{0}$ of $^{171}$Yb$^{+}$ in a magnetic field of 5.2 Gauss.

Collective single-qubit operations are preformed with co-propagating, $\sigma^+$-polarized Raman beams. Two qubit operations are
realized via the $\hat \sigma_{\phi}$ gate in the phase-sensitive geometry~\cite{Sorensen00,Lee05}. The ion chain is illuminated
with lin$\perp$lin-polarized Raman beams focused to a spot size $\approx$ 35 $\mu$m and propagating at a relative angle of 90 degrees, such that the
wavevector difference $\Delta \vec{k} \parallel \hat z$. The beatnote frequency of the two beams satisfies $\delta
\omega = \omega_{0} \pm \omega_{z,4} \pm \delta$, where $\omega_{z,4}$ is the highest frequency axial mode and $2\pi/\delta\approx t_{gate} = 30 \mu s$.

The detection is performed by state-dependent-resonance florescence, which returns photon counts proportional to the number of ions in the $\ket{1}$ (bright) state. For a detection duration of 400 $\mu$s, we detect an average of 9 photons from each ion in $\ket 1$ and 0.1
photons from each ion in $\ket 0$. Empirically, we find that the threshold measurement technique~\cite{Olmschenk07} results
in large measurement errors for 1 and 2 excited ion outcomes, which degrades the RB fitting. Instead, we perform detector tomography and process the RB data to reduce the measurement error~\cite{Keith18}. This allows us to reconstruct relative frequencies of the desired projection operators.

\begin{figure*} 
  \centering
    \includegraphics[width=1.0\textwidth]{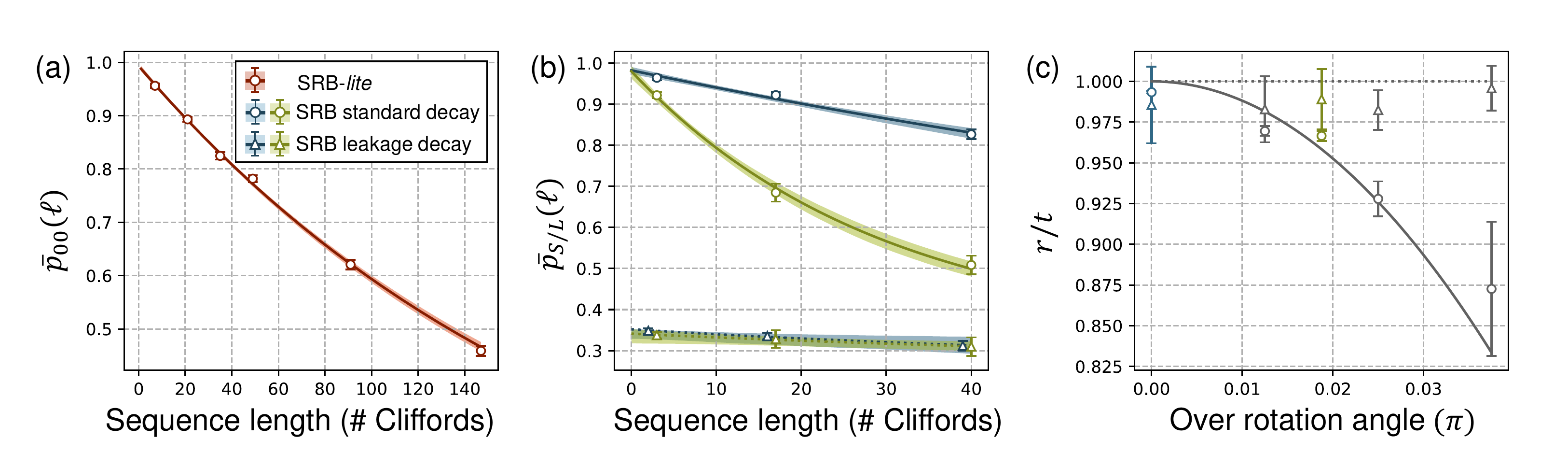}
  \caption{Experimental data from trapped-ion system. (a) SRB-\textit{lite} with estimated phase-gate errors of $1-f_{ZZ} =1.5(1) \times 10^{-3}$. The shaded region shows the one-sigma confidence interval of the decay fit from the semi-parametric bootstrap resampling. (b) Two example SRB experiments with standard analysis in circles with solid lines and leakage analysis in triangles with dashed lines: (blue) high-fidelity SRB with estimated phase gate errors of $1-f_{ZZ}' =2(1) \times 10^{-3}$, and (green) injected intensity error to show the change in the standard decay. The leakage points are offset in the x-axis for clarity. (c) SRB experiments with different magnitude intensity errors injected. The lines are from the analytic model discussed in App.~\ref{app:sims}. Circles with solid line show the standard decay rates for each magnitude error, while triangles with dashed line show the leakage decay rate for each magnitude error. The blue and green points from (b) are highlighted in the respective colors.}
\label{fig:high fid}
\end{figure*}

\section{Results and discussion} \label{sec:discussion}
We first performed SRB-\textit{lite} with sequences of qutrit-Clifford gates $[7,21,35,49,91,147]$ each with 33-36 randomizations (results plotted in Fig.~\ref{fig:high fid}a). Each sequence was repeated 100 times to build statistics. The points were fit to Eq.~\eqref{eq:rb_decay} with the least-squares method to give $r = 0.9949(3)$, which we used to estimate $f_{\textrm{Cliff}} = 0.9954(3)$ and phase-gate error rate of $1-f_{ZZ} = 1.5(1)\times 10^{-3}$.  Uncertainties correspond to one-sigma confidence intervals estimated by a semi-parametric bootstrap method~\cite{Meier06}. In this experiment, the qutrit-Cliffords were found with a slightly different method that resulted in an average of 3.056 phase-gates per qutrit-Clifford gate. The SRB-\textit{lite} method does not require detector tomography to reconstruct the relative frequencies needed. This makes the experiment simpler but does not identify the leakage error or reduce the number of fit parameters by fixing the $B$ parameter.

Next, we implemented SRB with sequences of qutrit-Clifford gates $[3, 17, 40]$ each with 50 random sequences (results plotted in Fig.~\ref{fig:high fid}b, solid lines with open markers). Each sequence was repeated 50 times to build statistics. For the leakage analysis, we used the even-parity outcomes. The results were fit to the decay curves derived in Sec.~\ref{sec:ssrb} with the least-squares method and plotted in Fig.~\ref{fig:high fid}a.
We found a standard decay of $r=0.9934(8)$ and a leakage decay of $t=0.985(23)$. These estimates translate to an extended-sub fidelity for the qutrit-Cliffords of $f_{\textrm{Cliff}}'=0.993(3)$ and a phase-gate error of $1-f_{ZZ}' = 2(1)\times 10^{-3}$. 

To validate that phase-gate errors dominate in SRB and SRB-\textit{lite}, we preformed standard single-qubit RB and measured a Clifford error of $1-F_{\textrm{Cliff}}=5.3(3) \times 10^{-5}$. We also note that the qutrit-Clifford gates have a very similar form to methods for generating arbitrary two-qubit unitaries, as demonstrated in Ref.~\cite{Hanneke10}, which makes the extended-sub fidelity for the qutrit-Cliffords a relevant metric as well. 

We also experimentally explored how SRB behaves when we inject errors into the system. We chose to inject over-rotation errors in the MS gate which translate to over-rotations in the phase gate plotted in Fig.~\ref{fig:high fid}c. For these tests, we used the same sequence lengths, reps, and trials per sequence as in Fig~\ref{fig:high fid}b. The standard decay $r$ decreases quadratically with the rotation magnitude while $t$ remains constant. The measured values match closely with the dashed lines, which are from a theoretical model described further in App.~\ref{app:sims}. The zero over rotation corresponds to the solid lines in Fig.~\ref{fig:high fid}b and the maximum over rotation value corresponds to the dashed line in Fig.~\ref{fig:high fid}b. Due to the close ion spacing we were not able to inject errors that cause leakage to the anti-symmetric subspace but we do study them numerically in App.~\ref{app:sims}. 

In certain cases, the leakage analysis may not be a reliable quantitative measure of leakage errors. This can happen if there are other leakage subspaces or there is not adequate control of the leakage subspace (see App.~\ref{app:ssrb}). These effects violate the assumptions made in the derivation of Eq.~\eqref{eq:leakage_decay} and cause other exponential decay terms to contribute to $\overline{p}_L(\ell)$. We also observe from the semi-parametric bootstrap of the numerical and experimental tests that the $t$ decay rate has much higher variance. This is due to a much shallower decay, which can make fitting difficult, and the large variance in the values of $\overline{p}_L(\ell)$. All of these issues will degrade the estimate of $t$.

However, the standard analysis is robust to all of these effects by including Weyl gates with the final inversion gate. This means the decay rate $r$ is still a valuable quantity and the leakage analysis could be used for qualitative information about leakage. The same analysis could be applied to any LB experiment to still extract information about gate fidelity in the presence of unknown or uncontrollable leakage.

\section{Generalized SRB}
SRB can be generalized to many other systems that are restricted to a non-universal set of gates. In this section we give a brief description of the general SRB procedure and outline the requirements.

In general, the set of available gates in a system is generated by a small set of interactions described by Hamiltonians $\{H_i\}$. We assume that each Hamiltonian can be implemented with arbitrary amplitude i.e. $U_i= \textrm{exp}[-i \theta_i H_i]$. These Hamiltonians form a representation of a Lie Algebra under commutation. If that algebra is equal to $\mathfrak{su}(d)$ for a $d$-dimenstional Hilbert space then the system is said to be controllable~\cite{Khaneja01} and any unitary in SU($d$) can be created with proper choices of $\{\theta_i\}$.

For SRB, we consider a set of Hamiltonians do not generate $\mathfrak{su}(d)$. Instead they generate a representation of another Lie algebra $\mathfrak{g}$, which is a sub algebra of $\mathfrak{su}(d)$. To perform SRB, we identify a subspace of the Hilbert space with dimension $d_1$ where $\mathfrak{g}$ forms $\mathfrak{su}(d_1)$. (If there are multiple such subspaces it is best to pick the largest one to capture the most information about the system.) The system is then said to be subspace controllable, and we can generate any unitary in that subspace. 

Identifying such a subspace may be a non-trivial task. For the trapped-ion example, we were aided by standard results on angular momentum addition in quantum mechanics, i.e. two spin $1/2$ systems add to form spin-1 (triplet) or spin-0 (singlet) states. Such relations could be applied to similar systems with arbitrary dimension subsystems. However, there may be extra leakage subspaces, which complicate the leakage analysis. In these cases, the leakage analysis will only provide identification of leakage rates and not quantitative measurement of their magnitude. Luckily, we can still extract some information on the quality of the gates from the standard decay analysis with the inversion procedure discussed in Sec.~\ref{sec:ssrb}. We consider this case in App.~\ref{sec:ssrb}. 

Once the required subspace is identified, we then generate a set of unitary gates that forms a two-design. For the procedure discussed in Sec.~\ref{sec:ssrb}, $d_1$ must be a prime or a power of a prime. This restriction is due to the construction of the Pauli or Weyl operators that are used to generate the Clifford group. If $d_1$ is not prime or a power of a prime, we could consider other unitary two-designs or randomize over Haar-random unitaries.

The selected gates are then generated by optimal control techniques~\cite{Machnes11}. Again, optimal control requires a numerical search with a full simulation of the unitary. This limits the $d_1$ to be small enough to simulate the corresponding unitary operations classically. While this is a limitation of SRB, standard RB is also limited to small system sizes (the largest published result is on three qubits~\cite{McKay19}). This is for two reasons: (1) the size of the Clifford group grows exponentially with the number of qubits and the optimal decomposition for each Clifford is found by exhaustive search, and (2) the difference in depth between Cliffords blows up with qubit dimension, which violates the standard RB assumption that each Clifford has the same error.

The remainder of the general procedure then follows standard RB and the trapped-ion example considered above. Random gates are applied for different sequence lengths and the total decay rate is measured. We can again determine the standard and leakage decays to derive the extended-sub fidelity.

\section{Conclusions}
We proposed a new method for benchmarking entangling qubit gates in trapped-ion systems, which is experimentally easier to implement and provides more information than standard methods used in such systems. While the method does not capture as much information as standard RB, it can be combined with LB to identify most errors that occur. We also proposed a modified LB procedure that allows for separating the leakage decay from other error rates that should be useful with any LB procedure.

We implemented this technique on a trapped-ion system to measure the quality of entangling gates. We verified a low error phase gate with $1-f_{ZZ}' = 2(1) \times 10^{-3}$. For our system, SRB is faster than both standard RB with two qubits (since it does not require transport) and Bell-state fidelity measurements (since it does not have standard quantum limit scaling with precision). This makes it an excellent method to use in the design-and-test cycle of tuning up high-fidelity entangling gates.

\begin{acknowledgments}
The authors would like to thank Patty Lee for her helpful comments on an earlier version of this manuscript.
\end{acknowledgments}

\appendix
\section{Python implementation}
A Python implementation of SRB-\textit{lite}, SRB, and detector tomography is also included as ancillary file to this submission. These methods are demonstrated in two Jupyter Notebooks that include simulation methods. They can also be edited to work with experimental implementation. Further details are given in the Jupyter Notebooks.

\section{Standard randomized benchmarking derivation} \label{app:standard_RB}
In this appendix, we describe standard RB with a final Pauli/Weyl gate included. As discussed in the main text, including this gate (1) catches total amplitude damping errors~\cite{Meier06}, (2) reduces the fitting parameters~\cite{Meier06, Harper19}, (3) allows one to show multiplicative precision~\cite{Harper19}, and (4) reduces the assumptions needed for LB. 

We work in the Liouville (or transfer matrix) representation of quantum processes. In this representation, density matrices and measurement operators are translated to supervectors $A \rightarrow | A \rrangle$, and processes are translated to superoperators $U [\cdot] U^{\dagger} \rightarrow \mathcal{U}$. The inner product of
two supervectors is equal to the trace overlap between the corresponding matrices $\llangle A | B \rrangle =
\textrm{Tr}(A^{\dagger} B)$. A supervector may be unnormalized, $\llangle A | A \rrangle = a$; for
example, a mixed state does not have a normalized supervector. However, at times we consider an orthonormal basis
of supervectors, that is a basis of orthonormal operators, $\llangle P _i | P_j \rrangle = \delta_{i,j}$ for a $d$-dimensional Hilbert space. The superoperators are related to the Kraus decomposition by
$\sum_i A_i \rho A_i^{\dagger} \rightarrow \sum_i A_i^* \otimes A_i | \rho \rrangle = \mathcal{A} | \rho \rrangle$. If
$\mathcal{A}$ is a trace preserving (TP) process, then $\llangle \mathds{1} | \mathcal{A} = \llangle \mathds{1} |$.

Each applied Clifford gate can be separated into its ideal action $\mathcal{C}_i$ and an error channel $\Lambda_i$. We also
include the effects of a state preparation error channel $\Lambda_P$ and a measurement error channel $\Lambda_M$. For zeroth-order RB we make the unrealistic assumption that $\Lambda_{i} = \Lambda$ for all $i$. Recent work
has shown that RB is robust to this assumption, although the final estimate will no longer have the same relation to the
process fidelity~\cite{Wallman18}.

The extra Pauli/Weyl gate $Q_k$ is compiled with the final inversion gate at the end of each sequence. We then study outcomes $m_k$ such that $\bra{m_k} Q_k = \bra{0}$, and define the survival probability as
\begin{equation} \label{eq:RB_dist}
p_{i, k}(\ell) = \llangle m_k | \Lambda_M \Lambda \mathcal{Q}_k \mathcal{C}_{i, \textrm{inv}} \cdots \Lambda
\mathcal{C}_{i,1} \Lambda_P |0 \rrangle,
\end{equation}
where $\mathcal{Q}_k = Q_k^* \otimes Q_k$,  $\ket{\psi}\bra{\psi} \rightarrow | \psi \rrangle$ for any pure state $\ket{\psi}$. The superopertors $\{ \mathcal{C}_i \}$ form a representation of the Clifford group under matrix
multiplication. Therefore, in each of the $i$ sequences we expand $\mathcal{C}_{i,j}$ in terms of relabeled elements $\{
\mathcal{D}_{i,j} \}$ defined as $\mathcal{C}_{i,1} = \mathcal{D}_{i,1}^{\dagger}$, $\mathcal{C}_{i,j} =
\mathcal{D}_{i,j}^{\dagger} \mathcal{D}_{i,j-1}$, and inversion gate $\mathcal{C}_{i,\textrm{inv}}=\mathcal{D}_{i,\ell}$.

Next, we average over all possible sequences $i$, which translates to an average over all Clifford gates at each position, yielding
\begin{equation} \label{eq:stand_surv}
\overline{p}_{k}(\ell) = \llangle m_k | \Lambda_M \Lambda \mathcal{Q}_k \left[ \tfrac{1}{N_C} \sum_i \mathcal{D}_i
\Lambda \mathcal{D}_i^{\dagger} \right]^{\ell} \Lambda_p | 0 \rrangle,
\end{equation}
where $N_C$ is the number of Clifford gates, and the position index $j$ in $\mathcal{D}_{i,j}$ is dropped.

Now, define the ``twirl'' $\mathcal{T}[\Lambda] =
\tfrac{1}{N_C} \sum_i \mathcal{D}_i \Lambda \mathcal{D}_i^{\dagger}$, which is a super-super operator, (an
operator acting on a superoperator). The twirl $\mathcal{T}$ maps any
process to the depolarizing process $\tilde{\Lambda}_d$~\cite{Magesan11}, which mixes an input state $\rho$ with the identity
$\tilde{\Lambda}_d[\rho] = r \rho + \tfrac{1-r}{d}\mathds{1}$. The parameter $r$ is the called the depolarizing parameter and relates directly to the process fidelity.
In the Liouville representation $\tilde{\Lambda}_d = r \mathbb{P} + | \mathds{1} \rrangle
\llangle \mathds{1} |$, where $| \mathds{1} \rrangle$ is the trace-normalized identity and $\mathbb{P}$ is the projection onto the orthogonal subspace $| \mathds{1} \rrangle
\llangle \mathds{1} |$. Substituting this superoperator into the previous equation gives
\begin{equation} 
\overline{p}_{k}(\ell) = A_k r^{\ell} + B_k,
\end{equation}
where $A_k = \llangle m_k | \Lambda_M \Lambda \mathcal{Q}_k \mathbb{P} \Lambda_p | 0 \rrangle$ and $B_k =\llangle m_k |
\Lambda_M \Lambda \mathcal{Q}_k | \mathds{1} \rrangle \llangle \mathds{1} | \Lambda_p | 0 \rrangle$.

Next, we average over all $d^2$ values of $Q_k$ and define $A = \tfrac{1}{d^2}\sum_k A_k$. The asymptote reduces
\begin{align}
\tfrac{1}{d^2} \sum_k B_k &= \tfrac{1}{d^2}\sum_k \llangle m_k | \Lambda_M \Lambda \mathcal{Q}_k| \mathds{1} \rrangle \llangle \mathds{1} | \Lambda_p |
0 \rrangle, \nonumber \\
&= \tfrac{1}{\sqrt{d}}\llangle \mathds{1} | \Lambda_M \Lambda | \mathds{1} \rrangle \llangle \mathds{1} | \Lambda_p | 0
\rrangle, \nonumber \\
&=\frac{1}{d},
\end{align}
where we used the relations by line number: ($1\rightarrow 2$) $\mathcal{Q}_k | \mathds{1} \rrangle \rightarrow Q_k \mathds{1} Q_k^{\dagger} = \mathds{1}$ and $\sum_k \llangle m_k | = d^{3/2} \llangle \mathds{1} |$ by the definition of a POVM, and ($2\rightarrow 3$) $\llangle
\mathds{1} | \Lambda = \llangle \mathds{1} |$ for any trace preserving map $\Lambda$ and $\llangle \mathds{1} | 0 \rrangle = d^{-1/2}$. We then get the fixed-asymptote
decay function
\begin{equation} \label{eq:standRB}
\overline{p}({\ell}) = A r^{\ell} + \tfrac{1}{d}.
\end{equation}
Subgroups of Pauli/Weyl gates, e.g. $\{\mathds{1}, X\}$, will also fix the asymptote~\cite{Harper19} but will require additional assumptions for the leakage analysis. 

We then estimate $\overline{p}({\ell})$ for different values of $\ell$ and fit the results to Eq.~\eqref{eq:standRB} to
estimate $A$ and $r$. The process fidelity of the depolarized process is equal to the process fidelity of the unknown
error channel~\cite{Nielsen02}.

In practice, we do not measure sequences with every possible final Pauli gate, but only sample from an even distribution over all final operators. This is similar to the standard RB practice of sampling from all sequences and, following similar arguments as in standard RB~\cite{Magesan12a}, we expect to approach the exact distribution with a reasonable number of samples.


\section{Derivation of subspace randomized benchmarking decay} \label{app:ssrb}
In this appendix, we repeat the previous derivation with leakage/seepage in the different erorr channels. We derive the decay curves in Eqs.~\eqref{eq:standard_decay} and ~\eqref{eq:leakage_decay}. We first follow similar steps as the derivation for leakage benchmarking given in Ref.~\cite{Wood18}. We then consider expansions of this derivation. 

\subsection{Basic decay derivation}
Let $d_1$ be the dimension of the RB subspace, $d_2$ be the dimension of the leakage subspace, and $d=d_1+d_2$. The projector on
the RB subspace is defined as $\mathds{1}_1$, with trace-normalized supervector $\tfrac{1}{\sqrt{d_1}} \mathds{1}_1 \rightarrow
| \mathds{1}_1 \rrangle$, and the projector onto the leakage subspace is defined as $\mathds{1}_2$, with trace-normalized supervector $\tfrac{1}{\sqrt{d_2}} \mathds{1}_2
\rightarrow | \mathds{1}_2 \rrangle$. 

Consider an ideal Clifford unitary $C$ acting on the RB subspace. Even in an ideal experiment, implementing
this Clifford will also implement a unitary process $V$ on the leakage subspace. Let $U = C \oplus V$ be the total unitary, or in the Liouville representation $\mathcal{U} = (C \oplus V)^* \otimes (C \oplus V)$. We expand this superoperator,
\begin{equation} \label{eq:exp_uni}
\mathcal{U} = \mathcal{C} + \mathcal{V} + \mathcal{A},
\end{equation}
where $\mathcal{C} = (C \oplus 0)^* \otimes (C \oplus 0)$, $\mathcal{V} = (0 \oplus V)^* \otimes (0 \oplus V)$, and
$\mathcal{A} = (C \oplus 0)^* \otimes (0 \oplus V) + (0 \oplus V)^* \otimes (C \oplus 0)$ (from now on we drop the direct sum
with zero notation e.g. $C \oplus 0 \rightarrow \mathcal{C}$ for brevity). The total superoperator $\mathcal{U}$ is block
diagonal since $\mathcal{C} \mathcal{V} = \mathcal{C} \mathcal{A} = \mathcal{A}
\mathcal{V} = 0$. However, the error channels $\Lambda$, $\Lambda_P$ and $\Lambda_M$ may not be block diagonal in this basis.
Some of these off-diagonal effects correspond to the leakage/seepage errors we wish to quantify.

Each SRB sequence $i$ consists of $\ell +1$ gates $\mathcal{U}_{i,j}$, which can be expanded as shown above. The survival probability is 
\begin{equation} \label{eq:RB_dist}
p_{i, k, m}(\ell) = \llangle m | \Lambda_M \Lambda  \mathcal{U}_{i, k, \textrm{inv}} \cdots \Lambda
\mathcal{U}_{i,1} \Lambda_P |0 \rrangle.
\end{equation}
There are a few differences between this survival probability and Eq.~\ref{eq:stand_surv}. For one, we do not fix the outcome $m$ with respect to the final Pauli/Weyl gate. This will later allow us to do the \textit{leakage analysis} discussed in the main text. Also, the measurement projectors $\{ \llangle m |\}$ are slightly more complicated since we now include the effects of the leakage subspace. The definition of POVMs  implies that some POVM elements will include projection onto the leakage subspace, $\sum_m \llangle m | = \sqrt{d} \llangle \mathds{1} |$ where $\mathds{1}$ is the identity on the entire system. For example, in the trapped-ion system the odd-parity measurement projects onto both the symmetric and anti-symmetric states. 

To simplify the derivation, Ref.~\cite{Wood18} makes the
following assumptions about the experiment: 
\begin{itemize}
\item \textit{Assumption 2}: $\mathcal{A}_{i,j} = 0$, which can be enforced by averaging over the relative phase between subspaces.
\item \textit{Assumption 3}: The ability to independently average over $\mathcal{C}_{i,j}$ and $\mathcal{V}_{i,j}$.
\item \textit{Assumption 4}: The action on the leakage subspace $\{ \mathcal{V}_{i,j} \}$ forms a unitary one-design.
\end{itemize}
These assumptions require some level of control over the leakage subspace, which is possible in the trapped-ion system, but may be difficult in other systems. For now, we make the same assumptions but consider their relaxation later. 

The result is that each gate is expanded $\mathcal{U}_{i,j} = \mathcal{C}_{i,j} + \mathcal{V}_{i,j}$ where $\mathcal{C}_{i,j}$ and $\mathcal{V}_{i,j}$ can be varied independently. As in standard RB, we expand each RB subspace superoporator $\mathcal{C}_{i,j} = \mathcal{D}_{i,j}^{\dagger} \mathcal{D}_{i,j-1}$, and additionally each leakage subspace superoperator $\mathcal{V}_{i,j} = \mathcal{W}_{i,j}^{\dagger} \mathcal{W}_{i,j-1}$.

The final gate becomes more complicated when considering the additional leakage subspace because it contains both an inversion and the final Pauli/Weyl gate. We associate any action on the leakage subspace with the inversion part since the action on the RB and leakage subspace can be varied independently. The inversion gate is selected to relate the sequence to a series of twirls $\mathcal{C}_{i, \textrm{inv}}=\mathcal{D}_{i, \ell}$, but in the leakage subspace there is an additional term when we expand $\mathcal{V}_{i,\textrm{inv}} = \mathcal{W}_{i,\ell+1}^{\dagger} \mathcal{W}_{i,\ell}$. The Pauli/Weyl gate is simplified under Assumption 2, $\mathcal{Q}_k = Q_k^* \otimes Q_k + \mathds{1}_2 \otimes \mathds{1}_2$. We combine this with the extra term on the leakage subspace $\mathcal{Q}_{k,i}'= Q_k^* \otimes Q_k +\mathcal{W}_i^{\dagger}$, where $i,\ell+1\rightarrow i$ since it is a random element from the set of leakage subspace gates. 

Now, we are ready to average over all possible sequences
\begin{widetext}
\begin{equation} \label{eq:pleak}
\begin{split}
\overline{p}_{k,m}(\ell) &= \tfrac{1}{N_C^{\ell}N_V^{\ell+1}} \sum_i \llangle m | \Lambda_M \Lambda 
\mathcal{Q}_k ( \mathcal{C}_{i,\textrm{inv}} + \mathcal{V}_{i,\textrm{inv}})\, \cdots \,
\Lambda ( \mathcal{C}_{i,1} + \mathcal{V}_{i,1}) \Lambda_P
|0 \rrangle,  \\
&= \llangle m | \Lambda_M \Lambda  \overline{\mathcal{Q}}_k \left[ \tfrac{1}{N_C N_V} \sum_i ( \mathcal{D}_{i} +\mathcal{W}_i) \Lambda ( \mathcal{D}_{i} + \mathcal{W}_{i}) \right]^{\ell}
\Lambda_P |0 \rrangle,
\end{split}
\end{equation}
\end{widetext}
where $N_V$ is the number different gates applied to the leakage subspace and $\overline{\mathcal{Q}}_k =\tfrac{1}{N_V} \sum_i \mathcal{Q}_{k,i}'=Q_k\otimes Q_k + |\mathds{1}_2 \rrangle \llangle \mathds{1}_2 |$.

For this analysis, we define a new twirl super-super operator and apply the one- and two-design definitions
\begin{align}
\mathcal{T}[\Lambda] &= \tfrac{1}{N_C N_V} \sum_i (
\mathcal{D}_{i} +\mathcal{W}_{i}) \Lambda ( \mathcal{D}_{i} + \mathcal{W}_{i})^{\dagger}, \nonumber \\
&= r \mathbb{P}_1 + t_1 | \mathds{1}_1 \rrangle \llangle \mathds{1}_1 | \nonumber \\
& + L | \mathds{1}_2 \rrangle \llangle \mathds{1}_1 | + S| \mathds{1}_1 \rrangle \llangle \mathds{1}_2 | + t_2 |
\mathds{1}_2 \rrangle \llangle \mathds{1}_2 |, \label{eq:Texp}
\end{align}
where $\mathbb{P}_1$ is the projection onto the part of the RB subspace orthogonal to $| \mathds{1}_1 \rrangle \llangle \mathds{1}_1|$. The coefficients are defined $r = \textrm{Tr}(\mathbb{P}_1 \Lambda)$, $L =\llangle \mathds{1}_2|\Lambda |\mathds{1}_1 \rrangle$, $S = \llangle
\mathds{1}_2| \Lambda |\mathds{1}_1 \rrangle$, $t_1=\llangle \mathds{1}_1| \Lambda |\mathds{1}_1
\rrangle=1-\sqrt{\tfrac{d_2}{d_1}}L$, $t_2 =\llangle \mathds{1}_2| \Lambda |\mathds{1}_2 \rrangle = 1-
\sqrt{\tfrac{d_1}{d_2}} S$. Our definitions of the leakage and seepage rates $L$ and $S$ differ slight from Ref.~\cite{Wood18} by factors of $\sqrt{d_1/d_2}$.

The next step is to solve for $\mathcal{T}[\Lambda]^{\ell}$ by diagonalizing $\mathcal{T}[\Lambda]$. The first term in
Eq.~\eqref{eq:Texp} is orthogonal to all other terms $\mathbb{P}_1 |\mathds{1}_i \rrangle \llangle
\mathds{1}_j| = 0$ for $i,j\in \{1,2\}$, but the remaining terms span a two-dimensional subspace $\{ | \mathds{1}_1 \rrangle, 
| \mathds{1}_2 \rrangle\}$ decomposed as
\begin{equation}
\begin{pmatrix} \label{eq:tp_mat}
1-\sqrt{\tfrac{d_2}{d_1}} L & S \\
L & 1-\sqrt{\tfrac{d_1}{d_2}} S
\end{pmatrix} 
= \lambda_+ \Pi_+ + \lambda_- \Pi_-,
\end{equation}
with eigenvalues $\lambda_{\pm}$ and eigenvectors $\Pi_{\pm}$,
\begin{align} \label{eq:pi_pm}
\lambda_+ &= 1, \nonumber \\
\lambda_- &= 1-\sqrt{\tfrac{d_2}{d_1}} L - \sqrt{\tfrac{d_1}{d_2}} S, \nonumber \\
\Pi_+ &= \frac{1}{d_2 L + d_1 S}
\begin{pmatrix} 
d_1 S & \sqrt{d_1 d_2} S\\
\sqrt{d_1 d_2} L & d_2 L 
\end{pmatrix}, \nonumber \\
\Pi_- &= \frac{1}{d_2 L +d_1 S}
\begin{pmatrix} 
d_2 L &  - \sqrt{d_1 d_2} S\\
-\sqrt{d_1 d_2} L & d_1 S
\end{pmatrix}.
\end{align}
Then $\mathcal{T}[\Lambda]^{\ell} = r^{\ell} \mathbb{P}_1 + \Pi_+ + \lambda_-^{\ell} \Pi_-$. In the main text, we defined $t=\lambda_-$.

We are now ready to combine our reductions to derive the survival probability with leakage. Substituting in the expressions
for $\mathcal{T}[\Lambda]^{\ell}$,
\begin{equation} \label{eq:p_k}
\begin{split}
\overline{p}_{k,m}(\ell) &= \llangle m| \Lambda_M \Lambda \overline{\mathcal{Q}}_k \mathbb{P} \Lambda_P | 0 \rrangle r^{\ell} \\
&+ \llangle m| \Lambda_M \Lambda \overline{\mathcal{Q}}_k \Pi_+ \Lambda_P | 0 \rrangle \\
&+ \llangle m| \Lambda_M \Lambda \overline{\mathcal{Q}}_k \Pi_- \Lambda_P | 0 \rrangle \lambda_-^{\ell} .
\end{split}
\end{equation}
We refer to the first term as the standard decay, the second as the asymptote, and the final term as the leakage decay.

To measure the standard decay (\textit{standard analysis} in the main text) we study the outcomes $m=m_k$ defined by the final Pauli/Weyl gate $\braket{m_k | Q_k |0} = 1$, and average over all $d_1^2$ Pauli/Weyl elements. This is the same approach as in App.~\ref{app:standard_RB}. The coefficient for the standard decay cannot be simplified further so define $A =\tfrac{1}{d_1^2} \sum_k \llangle m_k| \Lambda_M \Lambda \overline{\mathcal{Q}}_k \mathbb{P} \Lambda_P | 0 \rrangle$. We can reduce the asymptote in the following steps:
\begin{equation} \label{eq:b}
\begin{split}
B &= \tfrac{1}{d_1^2} \sum_k  \llangle m_k| \Lambda_M \Lambda \mathcal{Q}_k \Pi_+ \Lambda_P | 0 \rrangle, \\
&= \tfrac{\sqrt{d}}{d_1} \llangle  \mathds{1} | \Lambda_M \Lambda \Pi_+ \Lambda_P | 0 \rrangle, \\
&= \tfrac{\sqrt{d}}{d_1} \llangle \mathds{1} | 0 \rrangle  ,\nonumber \\
&= \tfrac{1}{d_1},
\end{split}
\end{equation}
where we used the relations by line number: (1$\rightarrow$2) $\overline{\mathcal{Q}}_k \Pi_+ = \Pi_+$ $\forall \, k$ and then the only $k$ dependence remains is in the outcomes $\sum_k \llangle m_k | =
\llangle \mathds{1}|d_1 \sqrt{d}$, (2$\rightarrow$3) $\llangle \mathds{1} | \Lambda_M \Lambda \Pi_+ \Lambda_P =
\llangle \mathds{1} |$ when $\Lambda_M$, $\Lambda$, and $\Lambda_P$ are TP on the full space and $\llangle \mathds{1} |
\Pi_+ = \llangle \mathds{1} |$ by inspection of Eq.~\eqref{eq:tp_mat}. A similar analysis can be applied to the leakage
decay by noting $\overline{\mathcal{Q}}_k \Pi_- = \Pi_-$ and $\llangle \mathds{1} | \Pi_- = 0$ to show that $\tfrac{1}{d_1^2}\sum_k \llangle m_k | \Lambda_M \Lambda \overline{\mathcal{Q}}_k \Pi_- \Lambda_P | 0 \rrangle =0$. The final decay is
\begin{equation}
\overline{p}_S(\ell) = A r^{\ell} + \tfrac{1}{d_1}
\end{equation}
This means that the method derived in App.~\ref{app:standard_RB} is robust to leakage/seepage errors.

To quantify leakage/seepage (\textit{leakage analysis} in the main text), we return to the distribution in Eq.~\eqref{eq:p_k} and
study the outcome $m=0$, which is chosen to have no projection onto leakage subspace. For example, in the trapped-ion system,
the even-parity outcomes are possible choices, whereas odd-parity outcomes have support in the anti-symmetric state. We then average over the Pauli/Weyl gates, which is equivalent to character benchmarking~\cite{Helsen19b}. The Pauli/Weyl operators form a unitary one-design and project onto the trivial irreducible representation $\Pi_Q = \tfrac{1}{d_1^2} \sum_k \overline{\mathcal{Q}}_k = | \mathds{1}_1 \rrangle \llangle \mathds{1}_1| + | \mathds{1}_2 \rrangle \llangle \mathds{1}_2|$, yielding
\begin{equation}\label{eq:p_0}
\begin{split}
\overline{p}_L({\ell}) &= \llangle 0| \Lambda_M \Lambda \Pi_Q \mathbb{P}_1 \Lambda_P | 0 \rrangle r^{\ell}  \\
&+ \llangle 0| \Lambda_M \Lambda \Pi_Q \Pi_+ \Lambda_P | 0 \rrangle  \\
&+ \llangle 0| \Lambda_M \Lambda \Pi_Q \Pi_- \Lambda_P | 0 \rrangle \lambda_-^{\ell} .
\end{split}
\end{equation}
However, in this case the standard decay term is equal to zero since $\Pi_Q \mathbb{P}_1 = 0$. 

The asymptote is easily calculated in the subspace spanned by $\{| \mathds{1}_1 \rrangle, | \mathds{1}_2 \rrangle\}$. The ideal projection
of the prepared state in this subspace is $\tfrac{1}{\sqrt{d_1}} |\mathds{1}_1\rrangle$. State preparation and measurement each may have leakage/seepage errors which cause them to overlap with the leakage subspace,
\begin{equation}
\begin{split}
\tfrac{1}{\sqrt{d_1}}\Lambda_P
|\mathds{1}_1\rrangle &= \tfrac{\sqrt{d_1}-\sqrt{d_2}}{d_1} L_P | \mathds{1}_1 \rrangle + \tfrac{1}{\sqrt{d_1}} L_P |
\mathds{1}_2 \rrangle, \\
\llangle \mathds{1}_1 | \Lambda_M \Lambda \tfrac{1}{\sqrt{d_1}} &=
\llangle \mathds{1}_1 |\tfrac{\sqrt{d_1}-\sqrt{d_2}}{\sqrt{d_1}} L_M +\llangle \mathds{1}_2 | \tfrac{1}{\sqrt{d_1}}
S_M
\end{split}
\end{equation}
where $L_P$ and $S_P$ are leakage and seepage rates for state preparation, respectively, and $L_M$ and $S_M$ are leakage and and seepage rates for measurement, respectively. 

The final step is to note that $\Pi_Q$ is the identity in this reduced subspace so $\Pi_Q \Pi_{\pm} = \Pi_{\pm}$. Then the asymptote is reduced to
\begin{equation}
\begin{split}
B &= \frac{d_1 S + \sqrt{d_1 d_2}(L S_m - S L_m)}{d_1(d_2 L + d_1 S)}, \\
 &= \frac{S}{d_2 L + d_1 S}+\epsilon_B,
\end{split}
\end{equation}
where $\epsilon_B = \mathcal{O}(L_M, S_M)$. The leakage decay coefficient is found similarly, but we omit the full definition for brevity and instead give the simplified form
\begin{equation}
C=\frac{(d_2/d_1) L}{d_1 S+d_2 L}  \lambda_- +\epsilon_C,
\end{equation}
 where $\epsilon_C = \mathcal{O}(L_M, S_M, L_P)$ and is related to SPAM errors. The net decay function is then,
\begin{equation}
\overline{p}_L({\ell}) = C \lambda_-^{\ell + 1} + B + \epsilon_B + \epsilon_C \lambda_-^{\ell}.
\end{equation}
In practice, we only fit the measured decay with the first two terms since the final two terms are likely much smaller
with high-fidelity state preparation and measurement. 

\subsection{Trapped-ion specific leakage detection}
In the trapped ion system $d_1=3$ and $d_2=1$, and Assumptions 3 and 4 are automatically enforced since $W_{i,j}$ 
trivially forms a unitary one-design $\mathcal{W}_{i,j} = | \mathds{1}_2 \rrangle \llangle \mathds{1}_2| \, \forall \, i,j$. 
Assumption 2 can be enforced by creating a phase-reversed set of qutrit-Cliffords gates by including one more phase gate (further details given in App.~\ref{app:control}). Averaging over
both the standard qutrit-Cliffords gates and the phase-reversed qutrit Cliffords then eliminates the $\mathcal{A}_{i,j}$ terms. Alternatively, in App.~\ref{app:no_reverse} we consider the derivation above but with Assumption 2 relaxed. 

Additionally, trapped-ion systems have more leakage sublevels that are not treated in the previous deriviation. For $^{171}Yb^{+}$ there are two other hyperfine levels per qubit ion that population may leak to due to sponetaneous emission. We consider such effects on the SRB derivation in App.~\ref{app:extra-sublevels}

\subsection{Averaging without the phase-reversed set} \label{app:no_reverse}
In the experiment, we chose not to average over the phase-reversed qutrit-Cliffords gates, discussed in the previous subsection, in order to fix the number of phase gates per Clifford. This choice is experimentally convenient since it means each Clifford has a very similar implementation. It also makes the experiment closer to the standard RB assumption that each Clifford has the same error channel. However, it means that Assumption 2 is not enforced. Ref.~\cite{Wood18} suggested that the randomization over the relative phase was not necessary and $\mathcal{A}_{i,j} \approx 0$. However, we find in this case that $\mathcal{A}_{i,j}$ is not small and in fact has some eigenvalues equal to one. This is an extreme case where the relative phase is the same for each Clifford gate since it is only changed by phase gates and each qutrit-Clifford gate is compiled with exactly three phase gates. 

We repeat the analysis for the trapped-ion system but without Assumption 2 to verify that the our procedure will still work. Assumptions 3 and 4 are still enforced since $\mathcal{W}_{i,j} = | \mathds{1}_2 \rrangle \llangle \mathds{1}_2| \, \forall \, i,j$.

We return to Eq.~\eqref{eq:pleak} but include $\mathcal{A}_{i,j}$. We expand the extra term $\mathcal{A}_{i, j}=\mathcal{B}_{i, j}^{\dagger} \mathcal{B}_{i, j-1}$ where $\mathcal{B}_{i,j} = D_{i,j}^* \otimes |\mathds{1}_2 \rrangle \llangle \mathds{1}_2| + |\mathds{1}_2 \rrangle \llangle \mathds{1}_2| \otimes D_{i,j}$. The final Pauli/Weyl term becomes $\mathcal{Q}_{k}' = Q_k^* \otimes Q_k + \mathcal{B}_{k}'^{\dagger} + |\mathds{1}_2 \rrangle \llangle \mathds{1}_2|$ where $\mathcal{B}_{k}'^{\dagger}=Q_{k}^* \otimes |\mathds{1}_2 \rrangle \llangle \mathds{1}_2| + |\mathds{1}_2 \rrangle \llangle \mathds{1}_2| \otimes Q_{k}$. Including these extra terms gives,
\begin{equation} \label{eq:expanded_twirl}
\begin{split}
\mathcal{T}[\Lambda] &=\tfrac{1}{N_C} \sum_i ( \mathcal{D}_{i} + \mathcal{W}_{i} +
 \mathcal{B}_{i}) \Lambda ( \mathcal{D}_{i} + \mathcal{W}_{i} + \mathcal{B}_{i})^{\dagger}   \\
&= \tfrac{1}{N_C} \sum_i \mathcal{D}_i \Lambda \mathcal{D}_i^{\dagger} + \mathcal{W}_i \Lambda \mathcal{D}_i^{\dagger} +\mathcal{D}_i \Lambda \mathcal{W}_i^{\dagger} + \mathcal{W}_i \Lambda \mathcal{W}_i^{\dagger}  \\
&+\mathcal{D}_i \Lambda \mathcal{B}_i^{\dagger} + \mathcal{W}_i \Lambda \mathcal{B}_i^{\dagger} + \mathcal{V}_i \Lambda \mathcal{D}_i^{\dagger} + \mathcal{B}_i \Lambda \mathcal{W}_i^{\dagger} + \mathcal{B}_i \Lambda \mathcal{B}_i^{\dagger},
\end{split}
\end{equation}
where the second line contains the terms we considered earlier and the third line contains the new terms. 

We numerically tested the effect of each term in Eq.~\ref{eq:expanded_twirl} by forming a matrix representation of the super-super operator, e.g. $\sum_i \mathcal{C}_i \Lambda \mathcal{C}_i^{\dagger} \rightarrow \sum_i \mathcal{C}^*_i \otimes \mathcal{C}_i | \Lambda )$ where $| \cdot )$ is the vectorized superoperator. Next, we take the eigendecomposition and analyze the magnitude of the eigenvalues. We see that all new terms except for the final term have small eigenvalues ($ \leq 0.1$). The small eigenvalues are exponentially suppressed in $\ell$ and we do not expect them to contribute to the decay curve. These are from an odd number of tensor products of $D_i$ unitaries, which average out to be small. The final term has an even number of $D_i$ unitaries, 
\begin{equation}
\begin{split}
\tfrac{1}{N_C} \sum_i  \mathcal{B}_i \Lambda \mathcal{B}_i^{\dagger} 
&= \tfrac{1}{N_C} \sum_i (D_i^* \otimes W_j) \Lambda (D_i^* \otimes W_i)^{\dagger}  \\
& +  (W_i^* \otimes D_i) \Lambda (W_i^* \otimes D_i)^{\dagger}  \\
& +  (D_i^* \otimes W_i) \Lambda (W_i^* \otimes D_i)^{\dagger}  \\
& +  (W_i^* \otimes D_i) \Lambda (D_i^* \otimes D_i)^{\dagger}.
\end{split}
\end{equation}
We see the final two terms have small eigenvalues (although slightly larger than before, $\leq 0.2$). The first two terms, however, each have a single eigenvalue equal to one. The corresponding eigenvectors are can be derived in a swapped basis,
\begin{equation}
\begin{split}
&\tfrac{1}{N_C} \sum_i  (D_i^* \otimes W_i)^* \otimes (D_i^* \otimes W_i) = \sum_{j,k} |k,a,k,a \rrangle \llangle j,a,j,a |,
\end{split}
\end{equation}
where $\ket{a} = \tfrac{1}{\sqrt{2}}(\ket{00} + \ket{11})$. The resulting operator projects the error channel $\sum_{j,k} \llangle k,a | \Lambda | k,a \rrangle  | j,a \rrangle \llangle j,a| = \textrm{Tr}(\Lambda \Pi_{\textrm{upper}}) \Pi_{\textrm{upper}}$ where $\Pi_{\textrm{upper}} = \sum_k  | k,a \rrangle \llangle k,a |$. A similar derivation can be applied to the second term giving  $\textrm{Tr}(\Lambda \Pi_{\textrm{lower}}) \Pi_{\textrm{lower}}$ where $\Pi_{\textrm{lower}} = \sum_k  | a,k \rrangle \llangle a,k|$. These two projectors correspond to parts of $\Lambda$ that map coherences between the RB and leakage subspace since $ |k,a \rrangle \rightarrow |k \rangle \langle a|$.

The Pauli/Weyl gate $\mathcal{Q}_k'$ has the same product with each of the previous projectors since it still acts as the identity over the $\{ | \mathds{1}_1 \rrangle$, $|\mathds{1}_2\rrangle \}$ subspace. The only change is that it has the additional $ \mathcal{B}_{k}'^{\dagger}$ term, which overlaps with $\Pi_{\textrm{upper}}$ and $\Pi_{\textrm{lower}}$. Further analysis of the extra term may show ways to eliminate these contributions.

The rest of the analysis is the same as above but with these two extra terms. The new terms do not contribute to the \textit{standard analysis} since $\llangle \mathds{1} | \Pi_{\textrm{upper/lower}} = 0$. Unfortunately, in the \textit{leakage analysis} the new terms persist. This means the leakage decay derived in Eq.~\eqref{eq:p_0} has two extra decay rates from $\Pi_{\textrm{upper/lower}}$. These terms do not relate to leakage/seepage errors and we cannot differentiate the effects. However, we expect that the state prepared and the measurements used have very little coherence between the two subspaces, which in practice likely limits the contribution of these two terms. This requires additional assumptions, but we believe they are reasonable at least in the trapped-ion case.

\subsection{Considering more leakage subspaces} \label{app:extra-sublevels}
For systems that have $n \geq 2$ leakage subspaces the analysis becomes more complicated, and we are not able to extract
the leakage decay exactly. For this sub-appendix we enforce Assumptions 2-4, which now require additional
control of the phase between each leakage subspace. Then $\mathcal{T}[\Lambda]$ can be reduced to a similar form as in
Eq.~\eqref{eq:Texp}, with the addition of leakage $L_{i,j}$ and seepage $S_{i,j}$ rates between the
$n+1$ subspaces,
\begin{equation} \label{eq:T2exp}
\begin{split}
\mathcal{T}[\Lambda] &= r \mathbb{P}_1 + \sum_{i=1}^{n+1} t_i | \mathds{1}_i \rrangle \llangle \mathds{1}_i | 
\\
& + \sum_{i<j}^{n+1} L_{i,j} | \mathds{1}_j \rrangle \llangle \mathds{1}_i | + \sum_{i>j}^{n+1} S_{i,j} | \mathds{1}_i \rrangle
\llangle \mathds{1}_j |.
\end{split}
\end{equation}

With $n$ leakage subspaces, we must consider a larger operator subspace spanned by $\{ | \mathds{1}_i \rrangle\}$ and diagonalize the corresponding $(n+1) \times (n+1)$ matrix. We can identify one eigenprojector under the assumption that $\Lambda$ is TP, which implies $\mathcal{T}[\Lambda]$ is TP. We label
this eigenprojector $\Pi_1 = | V_1 \rrangle \llangle \mathds{1} |$ where $V_1$ is a dual vector to $\llangle \mathds{1}|$,
and the corresponding eigenvalue is $\lambda_1 = 1$ since $\llangle \mathds{1} | \mathcal{T}[\Lambda] = \llangle \mathds{1} |$. (In the previous subsection this was $\Pi_+$.) The
remaining eigenvalues $\lambda_i$ for $i = 2,\dots,n+1$ are complicated functions of the leakage/seepage rates between
subspaces, and have corresponding eigenprojectors $\Pi_i$. We then substitute into Eq.~\ref{eq:pleak},
\begin{equation}\label{eq:pbig}
\begin{split}
\overline{p}_{k,m} ({\ell}) &= \llangle m| \Lambda_M \Lambda \overline{\mathcal{Q}}_k  \mathbb{P}_1 \Lambda_P | 0 \rrangle r^{\ell}  \\
&+ \llangle m| \Lambda_M \Lambda \overline{\mathcal{Q}}_k \Pi_1 \Lambda_P | 0 \rrangle  \\
&+ \sum_{i=2}^{n+1} \llangle m| \Lambda_M \Lambda \overline{\mathcal{Q}}_k  \Pi_i \Lambda_P | 0 \rrangle \lambda_i^{\ell} .
\end{split}
\end{equation}
The \textit{standard analysis} remains the same since $\llangle \mathds{1} | \Pi_{1} = \llangle \mathds{1} |$ and $\llangle \mathds{1} | \Pi_{i>1} = 0$. The asymptote corresponds to the known eigenvalue $\lambda_0$ and is again $B=1/d_1$.

Things do not reduce as nicely for the \textit{leakage analysis}. Eq.~\eqref{eq:pbig} has $n+1$ terms but, as in the two-subspace case, the standard decay is zero
since $\Pi_Q \mathbb{P}_1= 0$. The asymptote is nonzero and a function of the leakage/seepage
rates. There are additionally $n$ leakage decay parameters, which would require a multi-exponential fitting method to extract. Even if we use such a method, the different eigenvalues that define the leakage/seepage rates between subspaces could not be discerned.

\section{Qutrit subspace control} \label{app:control}
For the trapped-ion example considered, we generate the qutrit-Clifford gates based on a Cartan decomposition as shown in
Ref.~\cite{Khaneja01,Shauro2015}. In this appendix, we work with the spin-1 operators $J_x$, $J_y$, and $J_z$, as defined with respect to two-qubits in the main text. The first step in identifying the Cartan decomposition is to determine the maximum Abelian subalgebra of $\mathfrak{su}(3)$. This subalgebra is spanned by $J_z^2$ and $J_x^2$, which are generated by phase gates. Then any $U \in $SU(3) has an expansion,
\begin{equation}
U = K(\vec{n}_1, \phi_1) \exp[-i \theta_z J_z^2] \exp[-i \theta_x J_x^2]  K(\vec{n}_1, \phi_2),
\end{equation}
where $ K(\vec{n}_j, \phi_j)  = \exp[-i (\vec{n}_j \cdot \vec{J}) \phi_j]$ are SU(2) rotations performed with collective single-qubit gates. The
decomposition requires phase gates with arbitrary amplitude, which may be difficult to execute with high fidelity in practice.
Alternatively, we have found empirically that we can generate all qutrit-Clifford gates with the sequence
\begin{equation} \label{eq:control}
U = K(\vec{n}_1, \phi_1) U_{ZZ} K(\vec{n}_2, \phi_2) U_{ZZ} K(\vec{n}_3, \phi_3) U_{ZZ} K(\vec{n}_4, \phi_4),
\end{equation}
where $U_{ZZ} = \exp[-i J_Z^2 \pi/4]$. In practice, each arbitrary collective rotation is decomposed into one or two rotations around the $X-Y$ plane. We perform a numerical search~\cite{Machnes11} to determine the values of $\vec{n}_j$ and $\phi_j$ based on Eq.~\ref{eq:control}. 

We can also produce the phase-reversed qutrit-Clifford gate that puts the opposite phase between the subspaces thereby enforcing
Assumption 2 from App.~\ref{app:ssrb}. This is done by replacing one phase gate with two MS gates, which we call
an XX-YY gate $\exp[-i ZZ \pi/4] \rightarrow \exp[i XX \pi/4] \exp[i YY \pi/4]$, which is equivalent to iSWAP. The XX-YY gate has the same action on the 
RB subspace since $\exp[-i J_z^2 \pi/2] = \exp[i J_x^2 \pi/2] \exp[i J_y^2 \pi/2]$, but 
with a different relative phase on the leakage subspace. To generate a phase-reversed qutrit-Clifford group, we use the sequences found for the normal qutrit-Clifford gates but replace the final phase gate of each with the XX-YY gate.

\section{Bounds on fidelity} \label{app:fidelity}
The extended-sub fidelity given in Eq.~\eqref{eq:extnd_fid} provides a weak bound on the full fidelity. The full fidelity is the trace overlap between the error channel $\Lambda$ and the identity process $\mathds{1} \otimes \mathds{1}$, 
\begin{equation}
F = \tfrac{1}{d^2} \textrm{Tr}(\Lambda) = \tfrac{1}{d^2} \sum_i \llangle P_i | \Lambda | P_i \rrangle,
\end{equation}
for an orthonormal operator basis $\{ P_i \}$. Each term in the summation gives the change in $P_i$ due to the error channel $\Lambda$. Therefore, the full fidelity is the average change for a basis of operators.

In terms of SRB, the standard decay rate is equal to the trace overlap between $\mathbb{P}_1 = \mathds{1}_1 \otimes \mathds{1}_1 -|\mathds{1}_1 \rrangle \llangle \mathds{1} | = \sum_{i \in S} |P_i \rrangle \llangle P_i|$, where $S$ is an orthonormal basis off traceless operators acting on the RB subspace. Then $r = \tfrac{1}{d_1^2-1} \sum_{i \in S} \llangle P_i | \Lambda | P_i \rrangle$, which captures the average change of $d_1^2-1$ terms from the full fidelity. The leakage analysis measures $t+1 = \sqrt{\tfrac{d_2}{d_1}} L - \sqrt{\tfrac{d_1}{d_2}} S =  \llangle \mathds{1}_1 | \Lambda | \mathds{1}_1\rrangle + \llangle \mathds{1}_2 | \Lambda | \mathds{1}_2 \rrangle$. This contributes another two terms in the summation for full fidelity. The extended-sub fidelity is an average of these $d_1^2 +1 $ terms.

To weakly bound the full fidelity, we bound the magnitude of the other $d^2-d_1^2-1$ terms, which come from basis elements that mix population between the RB and leakage subspaces. These terms are each in the range $[-1,1]$, since the operator subspace is spanned by a set of traceless Hermitian operators with eigenvalues $\pm 1$. This gives the bounds,
\begin{equation}
\tfrac{d^2-d_1^2 +1}{d^2} (1 - f' ) \leq | F - f' | \leq \tfrac{d^2-d_1^2 +1}{d^2} (1+ f' ).
\end{equation}
The lower bound is proportional to the extended-sub infidelity, up to dimensional constants, which means it approaches zero for high-fidelity gates. However, the upper bound has a constant factor, which means there may be a process that has a high extended-sub fidelity but low full fidelity. 

The upper-bound is saturated when all other terms in the full fidelity are equal to -1. Consider the operator basis elements $P_{i,1} = |i_1 \rangle \langle i_2|$ and
$P_{i,2} =  |i_2 \rangle \langle i_1|$, where $\ket{i_1}$ are basis elements for the RB subspace and
$\ket{i_2}$ are basis elements for the leakage subspace. If $\llangle P_{i,1} | \Lambda | P_{i,2} \rrangle = -1$ for all $i$ then  $\Lambda$ generates a -1 relative phase between the RB and leakage subspace. This is the worst-case error we considered in Sec.~\ref{sec:sims} corresponding to SWAP errors. 

\section{Simulations with different error channels} \label{app:sims}
In this appendix, we compare analytic models of different error channels with simulations of SRB. We focus on the estimated quantities $r$, $t$, and $f_{ZZ}'$ to validate the expected behavior with the selected error channels.

\subsection{Analytic models}
While the extended-sub fidelity only weakly bounds the full fidelity, we find that for many typical error channels the values are close. In this section we directly calculate the SRB decay parameters and difference between the
extended-sub fidelity and full fidelity, for intensity errors, optical pumping, and inhomogeneous fields. We chose these error channels since they are parameterized by a single error magnitude and are easy to simulate, while still demonstrating different behaviors. The results of the analysis are shown in Table~\ref{tab:fid_comp}. Below is a brief description of each:

\begin{itemize}
\item \textit{Intensity errors}: Errors that cause the implemented phase gate to have a different amplitude than the expected value of $\pi/4$. This could be due to laser intensity fluctuations, Debye-Waller errors, or uneven intensity on the MS gate beams. We model these errors by applying an extra gate $U_e(\epsilon) = \exp[-i \epsilon ZZ]$ after each phase gate.

\item \textit{Optical pumping}: Errors that cause population to be pumped deterministically to one state (in our case modeled as $\ket{00}$). This could be due to stray light interacting with both ions. We model the errors as an amplitude damping channel with rate $\epsilon$.

\item \textit{Inhomogeneous fields}: Errors due to stray magnetic fields that fluctuate slowly with respect to the gate time. These errors could be due to trap electronics or other magnetic field sources in the lab. They are modeled as a phase-dampening channel with rate $\epsilon$.
\end{itemize}

\begin{figure*}
  \centering
    \includegraphics[width=\textwidth]{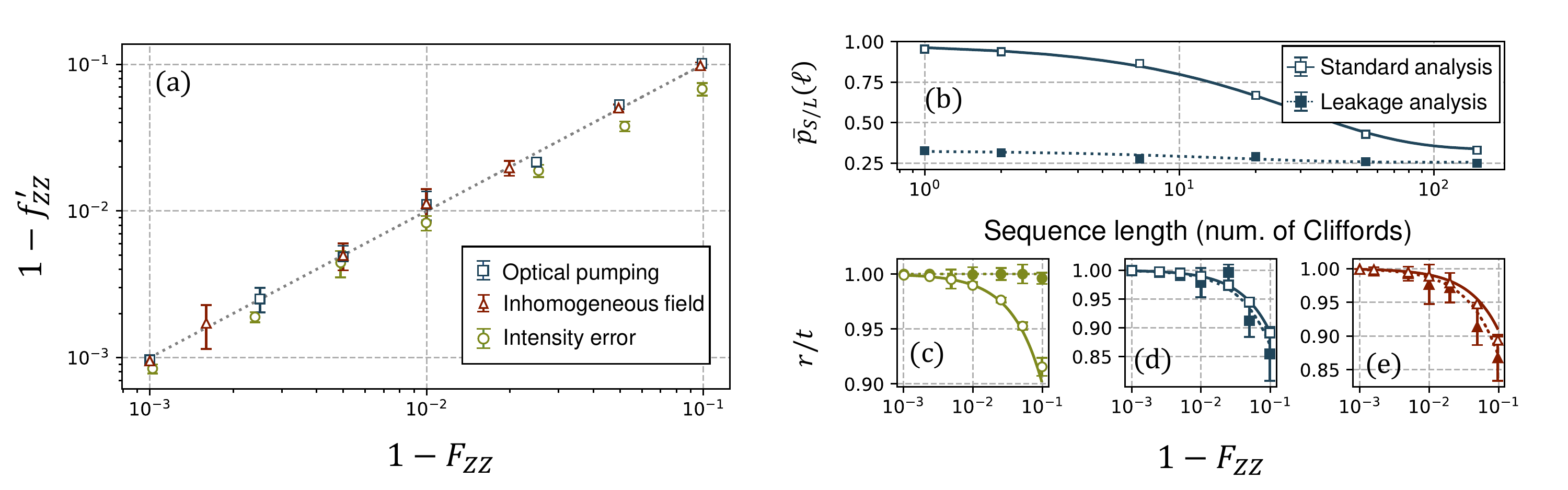}
  \caption{Simulations of several SRB experiments with three different error channels for the phase gate with various magnitudes. (a) Extended-sub infidelity from an SRB experiment vs full fidelity from analytic model with intensity errors (green circles), optical pumping errors (blue squares), and inhomogeneous fields (red triangles). The dashed line shows extended-sub fidelity equal to full fidelity. (b) Example results from an SRB experiment with standard decay (solid line with empty squares), and leakage decay (dashed line with filled squares). The points represent simulated data and the lines are fits to Eq.\eqref{eq:standard_decay} and~\eqref{eq:leakage_decay}. (c-d) Scaling of standard decay $r$ (open markers) and leakage decay $t$ (filled markers) with infidelity for the three error channels simulated. The points are from the SRB simulations and the lines are from the analytic model.}
 \label{fig:3error_sim}
\end{figure*}

\subsection{Simulations}
We studied SRB's performance with the three error channels described above. The results are plotted in Fig.~\ref{fig:3error_sim}. We simulated SRB experiments with injected errors in the phase gate with $1-F \in [10^{-3},10^{-1}]$, single-qubit depolarizing errors of magnitude $10^{-5}$ per gate, and depolarizing SPAM errors of magnitude $10^{-3}$. For most values of $1-F$ the sequence lengths are exponentially distributed up to $1/(1-F)$. For larger values of $1-F$, we used the sequence lengths $1,2,...,10$, which perform better. For each sequence length, we generated 100 random sequences of gates and sampled each of those sequences 100 times to build statistics. We simulated the trapped-ion parity measurement, which detects the number of ions in the bright state by state-dependent-florescence, and performed detector tomography to better differentiate the outcomes. For the leakage analysis, we used the even-parity outcomes. Uncertainties are calculated by the semi-parametric bootstrap method~\cite{Meier06} with one-sigma confidence intervals plotted. An example of a single SRB experiment is plotted in Fig.~\ref{fig:3error_sim}b with fits to the standard decay model (solid line) and leakage decay model (dashed line). 

Fig.~\ref{fig:3error_sim}a shows the full infidelity vs. the estimated extended-sub fidelity of the phase gate from SRB. For the errors simulated, we see that the extended-sub infidelity estimated by SRB is very close to the full infidelity. 

Figs.~\ref{fig:3error_sim}c-d, shows the extracted standard and leakage decay rate from each experiment with (c) intensity errors, (d) optical pumping, and (e) inhomogeneous fields. We calculated the expected values of $r_{ZZ}$ (solid lines) and $t_{ZZ}$ (dashed lines) based on the error models described above. We then scaled the measured values $\hat{r}$ and $\hat{t}$ from SRB assuming that the phase errors dominate and are depolarizing $\hat{r}_{ZZ} = \hat{r}^{1/3}$ and $\hat{t}_{ZZ} = \hat{t}^{1/3}$. This assumption is clearly not true since we inject error channels that are not depolarizing, but it allows us to relate the models for phase gate errors directly to SRB measurements. We observe that the values of $\hat{r}_{ZZ}$ and $\hat{t}_{ZZ}$ are similar to that predicted from the analytic model described above. The intensity error has $\hat{r}_{ZZ}$ decay but no $\hat{t}_{ZZ}$ decay while the optical pumping and inhomogeneous magnetic field errors have visible decay for both. 

For larger error magnitudes the value of $\hat{t}_{ZZ}$ differs more from the expected value. This is due to fitting troubles in the leakage analysis that are discussed more in Sec.~\ref{sec:discussion} but the analysis still gives an indication of leakage errors.

\begin{table}
\begin{tabular}{|l|c|c|c|}
\hline
\textbf{Error channel} $\quad\quad\quad\:$  &  $\quad \quad \boldsymbol{r} \quad \quad$ & $\quad \quad \boldsymbol{t}\quad \quad$ & $\, \boldsymbol{|F_{ZZ} - f_{ZZ}'|} \,$  \\ \hline
Intensity error    &   $1-\epsilon^2$  & 1 &  $\epsilon^2/5$  \\ \hline
Optical pumping      & $1-13 \epsilon/12$ &  $1-4 \epsilon/3$  &  $2 \epsilon/5$  \\ \hline
Inhomogeneous field & $1-13 \epsilon/6$ & $1-8 \epsilon/3$ & $\epsilon^2/40$ \\ \hline
\end{tabular}
\caption{Lowest order expansions of standard decay $r$, and leakage decay $t$, and difference between fidelities $|F_{ZZ} - f_{ZZ}'|$ for a few selected error channels. The error channels are each specified by a single error magnitude parameter $\epsilon$.}
 \label{tab:fid_comp}
\end{table}

\section{Phase-gate error channels} \label{app:errors}
In this appendix we categorize some common error channels in the phase gate according to their action on the
RB (symmetric) and leakage (anti-symmetric) subspaces. To determine if the error channels are subspace preserving (type-1) or leakage inducing (type-2) we look
at the leakage and seepage rates defined in the App.~\ref{app:ssrb}. When $L = S = 0$ then
the error is type-1 otherwise the errors are type-2.

The phase gate is created by applying wrapper pulses before and after an MS gate, and, therefore, most errors we study are MS gate errors. Let $U_{MS}$ be the unitary that describes the ideal MS gate. Within the Lamb-Dicke approximation and the resolved side-band limit, the propagator has a
simple closed form that can be found by considering the Magnus expansion $U_{MS}\approx\textrm{exp}\lbrack-i\int dt'
H(t')-\frac{1}{2}\int\int dt'dt''\lbrack H(t'),H(t'')\rbrack\rbrack$. The first term in this expression is usually
referred
to as the spin-dependent force and the second is referred to as the spin-dependent phase. Ideally, the spin-dependent
force
integrates to zero and one is left with
$U_{MS}=\textrm{exp}[-i\hat{S}^2(\frac{\eta\Omega}{2\delta})^2(\delta\tau_g-\textrm{sin}(\delta\tau_g))]$ where
$\hat{S}$ is
the collective spin operator, $\eta$ is the Lamb-Dicke parameter, $\Omega$ is the Rabi frequency, $\delta$ is the
detuning from resonance and $\tau_g$ is the gate time.
\\[8pt]
\noindent
\textit{Intensity/Lamb-Dicke/Debye-Waller errors}: Fluctuations in the Rabi frequency $\Omega$ can usually be attributed to either laser intensity noise, the Debye-Waller effect, or the Lamb-Dicke nonlinearity~\cite{Sorensen00}. These errors re-scale the rotation angle of the spin-dependent phase, but the operator $\hat{S}^2$ remains symmetric under qubit exchange, implying that these errors are subspace preserving.
\\[8pt]
\noindent
\textit{Trap frequency error}: Trap frequency fluctuations are another common error source~\cite{Hayes12}, and can be thought of as fluctuations in the detuning parameter $\delta$. While these errors manifest themselves as fluctuations in the spin-dependent phase, they also alter the structure of the spin-dependent force component. Ideally the spin-dependent force component is given by $\textrm{exp}\lbrack-i\int dt' H(t')\rbrack=\textrm{exp}\lbrack\hat{S}(\alpha^*(t)a^{\dagger}-\alpha(t)a)\rbrack$ and $\alpha(t_{gate})\rightarrow 0$. However, trap frequency fluctuations will lead to a non-zero $\alpha$ and $\hat{S}$ is generally not symmetric with respect to qubit exchange~\cite{Lee05}. In order for $\hat{S}$ to be symmetric under qubit exchange,
the laser phase on the two different ions must be the same as their relative motional phase in the motional eigenmode used
for the gate. For example, if one opts to use the center-of-mass mode for the gate operation, $\hat{S}$ is only symmetric if
the ion-spacing $x_0$ satisfies $k x_0=m\pi$ where $m$ is even with $k$ being the laser's wave-vector component that is
parallel to the relevant principle axis of motion. If one chooses to use a motional mode where the two ions' motion is
out-of-phase, then $m$ is required to be an odd integer. Therefore trap frequency errors can be either subspace preserving or leakage inducing.
\\[8pt]
\noindent
\textit{Spectator mode coupling}: Spectator mode coupling is close in structure to trap frequency fluctuation errors. The inclusion of
spectator modes into the propagator changes the amount of spin-dependent phase that is accumulated, but also introduces additional
spin-dependent forces. In the case of a 4-ion crystal, no matter which motional mode is used for the gate, there will
always
be spectator modes of both types, those where the qubit ions' motion is in-phase and those where it is out-of-phase.
This
implies that no matter what the ion-spacing is, there will be residual spin-dependent force operators that are leakage inducing.
\\[8pt]
\noindent
\textit{Carrier coupling}: Another off-resonant transition to be considered is that of the so-called carrier transition, which changes the
spin without changing the motional state. As discussed in~\cite{Sorensen00}, this error channel can be made to vanish if the
gate
time is chosen to be commensurate with the mode frequency, but will be non-zero in the presence of small timing errors.
Unlike spectator-mode transitions, this error will introduce operators whose exchange symmetry only depends on the laser phase
at the location of the ions and is subspace preserving when $m$ is even. Therefore, carrier coupling can be either subspace preserving or leakage inducing.
\\[8pt]
\noindent
\textit{Carrier frequency fluctuations}: Fluctuations of the carrier laser frequency are
equivalent to collective qubit frequency fluctuations. In general, qubit frequency fluctuations are leakage inducing if there are differential shifts on different qubits. However, in the case where the two data ions are only $~2.5\mu m$ apart, common-mode field fluctuations should dominate,
and therefore are subspace preserving to a good approximation. 
\\[8pt]
\noindent
\textit{Uneven illuminations}: Counter-intuitively, uneven illumination of the two ions is also subspace preserving. Under uneven ion illumination, the collective spin-operator gets perturbed as
$\hat{S}=\hat{s}_1+\hat{s}_2\rightarrow(1+\delta\Omega/2)\hat{s}_1+(1-\delta\Omega/2)\hat{s}_2$, which is not symmetric under qubit exchange. However, the spin-dependent force still integrates to zero in this case and the square of this operator is still symmetric under qubit exchange, meaning that this error is subspace preserving.
\\[8pt]
\noindent
\textit{Heating}: Heating of the motional mode can be thought of as instantaneous momentum kicks randomized in time. These momentum kicks do not prevent the phase space loops from closing, rather they only change the amount of area that is traced out. Therefore, this error channel is subspace preserving.
\\[8pt]
\noindent
\textit{Spontaneous emission}: Spontaneous emission drives population outside the computational subspace and is therefore leakage inducing.
\\[8pt]
In order to generate Table~\ref{tab:table1} which summarizes which phase gate errors are subspace preserving, we go through the same analysis for the wrapper pulses and then XOR those results with those from the MS gate. In our case, where the wrapper pulses are collectively applied, the only single-qubit gate errors that couple the subspaces come from uneven illumination of the two qubits, which makes this error leakage inducing.

\end{document}